\documentclass[prb,aps,twocolumn,epsfig,eqsecnum]{revtex4}
\usepackage{epsfig}
\usepackage{wasysym}
\usepackage{bm}
\usepackage{pstricks}

\begin{document}
\title{Spectral properties in the charge density wave phase of the half-filled
Falicov-Kimball Model}
\author{S.R.Hassan$^{1,2}$, H. R. Krishnamurthy$^{2,*}$}
\affiliation{$^1$D\'{e}partement de Physique,Universit\'e de Sherbrooke,
Qu\'ebec, Canada J1K 2R1}
\affiliation{$^2$Centre for Condensed Matter Theory, Department of Physics,
Indian Institute of Science, Bangalore 560 012, India}
\begin{abstract}
We study the spectral properties of charge density wave (CDW) phase
of the half-filled spinless Falicov-Kimball model within the
framework of the Dynamical Mean Field Theory. We present detailed
results for the spectral function in the CDW phase as function of
temperature and $U$. We show how the proximity of the non-fermi
liquid phase affects the CDW phase, and show that there is a region
in the phase diagram where we get a CDW phase without a gap in the
spectral function. This is a radical deviation from the mean-field
prediction where the gap is proportional to the order parameter.
\end{abstract}

\maketitle

\section{Introduction}
Metzner and Vollhardt \cite{volhardt_2} pioneered a new approach to
the study of strongly  correlated electron systems which is exact in
the limit of infinite dimensionality. Generally referred to as
Dynamical Mean-Field Theory (DMFT), the method has been developed
further in the subsequent years and has led to substantial progress
in our understanding of these systems\cite{Jarrell_2,Georges_2}.
Soon after the work of Metzner and Vollhardt\cite{volhardt_2}, in a
series of papers\cite{Brandt_2}, Brandt and Mielsch  showed that the
large dimensional limit of the spin-less Falicov-Kimball
model (SFKM)\cite{fkm} is exactly soluble and studied various aspects
of the solution. One of the first issues examined in the SFKM using
the DMFT was that of ordering into a two sublattice,
``checkerboard'', charge density wave (CDW) state that is expected
to arise in the model at half-filling on a hyper cubic lattice.
Because the hyper cubic lattice is bipartite, a transition at a
non-zero temperature is expected\cite{Kennedy_2} for all $U$  and
indeed this is true within the DMFT as well\cite{Brandt_2}.

Since the original work\cite{Brandt_2}, a number of other studies of
the `uniform phase' at high temperatures, the CDW order and of phase
separation away from half filling have followed
\cite{Si_2,Ruckenstien_2,Dong1,Freerick1,Gruber_2,Letfulov_2}, on
both the hyper cubic and Bethe lattices. The homogeneous or uniform
phase at high temperatures is a non-Fermi liquid\cite{Si_2}. A
detailed study of transport in this phase has been carried out by
Moller {\it et al}\cite{Ruckenstien_2}. Van dongen\cite{Dong1}
pointed out that for a lattice with a bounded density of states
(DOS) the half-filled system in the uniform phase displays a
metal-insulator transition at a non-zero U. Since the DMFT is in the
thermodynamic limit, one can determine the transition temperature
for the continuous transition from the uniform to the CDW phase
simply by finding the temperature at which the charge susceptibility
at wave-vector ${\vec Q}_0 \equiv (\pi, \pi,...)$ diverges. Thus one
can obtain the phase diagram of the half filled system purely from
the knowledge of the uniform solution. The phase diagram for
commensurate and incommensurate filling (away from half filling) in
one and two dimensions within the frame work of the DMFT has been
studied by Freericks\cite{Freerick1}.

Despite all this work, the properties of the CDW phase itself do not
seem to have been explored a great deal. In this contribution, we
study the spectral and transport properties of the charge density
wave (CDW) phase of the half-filled SFKM in the DMFT approximation
at a level of detail not reported before, to the best of our
knowledge. We present detailed results for the spectral functions in
the CDW phase as functions of temperature and U. We show how the
non-Fermi liquid behavior in the uniform phase affects the CDW
phase, and show that there is a region in the phase diagram, not
emphasized in the literature before, where we get a CDW phase
without a gap in the spectral function. This is a radical deviation
from the static mean-field prediction where the gap is proportional
to the order parameter, and the CDW phase is always gapped. The
`transition temperature' $T_l(U)$ between this novel gapless `phase'
and the conventional gapped CDW phase has an interesting `reentrant'
structure as U increases.

\section{DMFT of SFKM; Formalism}
%\noindent

The Hamiltonian for the SFKM  on a lattice with sites labeled by $i$
can be written as
\begin{eqnarray}
H  &  = & -\sum_{i,j}t_{ij}b_{i}^{\dag}b_{j} +
\epsilon_{\ell}\sum_{i}\ell_{i}^{\dag}\ell_{i} \nonumber\\ & &
-\mu\sum_{i}(b_{i}^{\dag}b_{i} + \ell_{i}^{\dag}\ell_{i}) + U
\sum_{i}b_{i}^{\dag}b_{i}\ell_{i}^{\dag}\ell_{i} \mbox{.}
\end{eqnarray}
It includes spin-less conduction or band electrons $b$ with hopping
parameters $t_{ij}$, which, in this paper, we assume to be non zero
only for nearest neighbors, and localized $\ell$ electrons conserved
at each site with site energy $\epsilon_{\ell}$. $\mu$  is the
chemical potential for the  $b$ and $\ell$ electrons. There is an
on-site coulomb repulsion $U$ between $b$ and $\ell$. In this paper,
we confine ourselves to the particle hole symmetric case, with
$N^{-1} \sum_{i}{\bar n}_{\ell i} \equiv N^{-1} \sum_{i} < n_{\ell
i}> = 1/2$ {\it and} $N^{-1} \sum_{i}{\bar n}_{bi} \equiv N^{-1}
\sum_{i} < n_{bi}> = 1/2$ where $N$ is the total number of sites in
the lattice. On bipartite lattices this is achieved by the choice
$\epsilon_{\ell}=0$, $\mu=\frac{U}{2}$.

The exact solubility of the SFKM within the DMFT approximation
arises as follows\cite{Brandt_2}. In infinite dimensions, or in
finite dimensions within the DMFT, the self energy, irreducible
vertex functions etc., are purely local, and in diagrammatic
perturbation theory, for example, are given by sums of skeleton
graphs involving only the local Green's function. The problem is
hence mapped to a single site or impurity problem embedded in a self
consistent medium representing all the other sites of the lattice.
In a functional integral formalism, the resulting single site
effective action at site $i$ is given by\cite{Jarrell_2,Georges_2}

\begin{eqnarray}
{\cal S}_{eff} &=& -\int_{0}^{\beta} \int_{0}^{\beta} d\tau d\tau'
b_{i}^{\dag}(\tau) {{\cal G}^{ii}}^{-1}(\tau - \tau') b_{i}(\tau')
- \beta \mu n_{\ell i}\nonumber \\
+& & U n_{\ell i}\int^\beta_0 d \tau b_{i}^{\dag}(\tau)b_{i}(\tau)
\label{eq-seff}
\end{eqnarray}

Here $b_{i}^{\dag}(\tau)$ and $b_{i}(\tau)$ are fluctuating
fermionic Grassmann fields corresponding to coherent states of the
$b$ electrons, and $n_{\ell i}=$ 1 or 0 corresponding to the $\ell$
state at site $i$ being occupied or not. ${\cal G}^{ii}(\tau)$ is
the bare on site local propagator (also referred to as the host or
medium propagator) for the self consistent single site problem. It
is essentially the local propagator at site $i$ excluding the self
energy processes at that site (since they will be recovered by
solving the single site problem) but including the self energy
processes at all the other sites of the lattice in some average way.
(Site indices are being kept track of in the equations above with a
view to using them in contexts involving broken symmetry solutions
such as the CDW phase.)

Since the effective action eq. (\ref{eq-seff}) is quadratic in the
variable $b$ for a fixed $n_{i\ell}$, the local Green's function can
be calculated exactly for an arbitrary host Green's function as
\begin{equation}
G_{n}^{ii} = \frac {\mbox{w}_{0i}} { ({\cal G}_{n}^{ii})^{-1}}+ \frac
{\mbox{w}_{1i}} { ({\cal G}_{n}^{ii})^{-1}-U } \,. \label{localG}
\end{equation}
Here $G_{n}^{ii} \equiv G^{ii}({i\omega_n})$ (and similarly for
${\cal G}$, etc.), $ {i\omega_n} \equiv i(2n+1)\pi T$ are the
Fermionic Matsubara frequencies, and w$_{1i}$ and w$_{0i}$ = (1
-w$_{1i}) $ are the annealed probabilities for the $\ell$ state at
site $i$ being occupied and empty respectively. w$_{1i}$, also
equal to the average $\ell$ electron number at site $i$, can be
calculated as
\begin{equation}
\mbox{w}_{1i} = {\bar n}_{\ell i} = \frac{Z_{1i}}{Z_{0i}+Z_{1i}}\mbox{,}
\end{equation}
in terms of $Z_{1i}$ and $Z_{0i}$, the constrained partition
functions obtained by summing $\exp(-S_{eff})$ over  all the
fluctuating $b$ configurations for fixed $n_{\ell i}=$ 1 or 0
respectively. It is straightforward to verify that
\begin{equation}
Z_{\nu i} = \exp \left[ \beta \mu \delta _{\nu 1} + \sum_{n} \ln
\left(   U \delta _{\nu 1} - ({\cal G}_{n}^{ii})^{-1} \right)e^{i
\omega_n 0^+} \right] \mbox{.} \label{Znu}
\end{equation}

A renormalized or effective $\ell$ electron energy at site $i$ can
be defined by expressing w$_{1i}$ as
$n_{F}^{-}(\epsilon^*_{{\ell}_i}-\mu$), whence
\begin{equation}
\epsilon^*_{{\ell}_i}=\mu+Tln(\frac{Z_{0i}}{Z_{1i}}).
\label{effl}
\end{equation}

The local Green's function $ G_{n}^{ii}$, host Green's function
${\cal G}_{n}^{ii} $ and self energy $ \Sigma_{n}^{ii}$ are related
by the local Dyson equation,
\begin{equation}
(G_{n}^{ii})^{-1} = ({\cal G}_{n}^{ii})^{-1} -
\Sigma_{n}^{ii}\mbox{.} \label{Localdyson}
\end{equation}
Using this equation in reverse, in the form,
\begin{equation}
({\cal G}_{n}^{ii})^{-1} = \Sigma_{n}^{ii} + ({G_{n}^{ii}})^{-1}
\mbox{,} \label{ld2}
\end{equation}
substituting into eq. (\ref{localG}), and cross-multiplying, one
gets\cite{Brandt_2} a quadratic equation which can be solved to
express the self energy directly in terms of the local Green's
function and $w_{1i} $. The result is
\begin{equation}
\Sigma_{n}^{ii}=\frac{U}{2}-\frac{1}{2G_{n}^{ii}}\left(  1 -
\sqrt{1 + (UG_{n}^{ii})^{2} + 4 (\Delta \mbox{w}_{1i})
UG_{n}^{ii}}\right) \mbox{,} \label{Sigmaii}
\end{equation}
where we have introduced
\begin{equation}
\Delta \mbox{w}_{i} \equiv (\mbox{w}_{1i}-\frac{1}{2}) = (\frac{1}{2} - \mbox{w}_{0i})
\mbox{.}
\end{equation}
(The sign of the root chosen in writing eq. (\ref{Sigmaii}) is so as
to reproduce the known analyticity properties of the self energy.)
In terms of the ratio $x_i \equiv Z_{1i}/Z_{0i}$ which can be
calculated, using eq. (\ref{Znu}), as
\begin{equation}
x_i = \exp \left[ \beta \mu + \sum_{n} \ln \left( \frac {({\cal
G}^{ii}_{n})^{-1} - U } {({\cal G}_ {n}^{ii})^{-1}} \right)
e^{i\omega_{n}0^{+}} \right]\mbox{,}
\end{equation}
we can write
\begin{equation}
\Delta \mbox{w}_{i} = \frac{1}{2} \left(\frac{x_{i} - 1}{x_{i} +
1}\right)\mbox{.}
\end{equation}
We note that, although according to eq. (\ref{Sigmaii})
$\Sigma_{n}^{ii} $ seems to depend explicitly only on $G_{n}^{ii}$
at the same Matsubara frequency, implicitly, via its dependence on
$\Delta w_{i}$ and hence on $x_i$, it is in general a functional of
$G_{m}^{ii}$ at  all frequencies $i \omega_{m}$.

To complete the DMFT procedure, one has to next invoke the DMFT self
consistency relation which determines the local Green's functions in
terms of the local self energies via lattice propagators, based on
treating the local self energy as a good approximation (which is
exact in infinite dimensions) to the lattice self energy. For this
purpose we need to use the lattice Dyson equation,
\begin{equation}
(i\omega_{n} +\mu
-\Sigma_{n}^{ii})G_{n}^{ij}-\sum_{l}t_{il}G_{n}^{lj}=\delta_{ij}\mbox{.}
\label{latdyson}
\end{equation}
The properties of this relation depend very much on the phase one
is interested in.

\section{The uniform phase and the CDW instability }
%\noindent

At high temperatures the $b$ electrons and $\ell$ electrons  are
uniformly distributed on the lattice for all electron
concentrations; there is no long range order, and no broken
symmetry. In this case of a homogeneous or uniform phase,
w$_{1i}$=w$_{1}$, $\Sigma_{n}^{ii}=\Sigma_{n}$, etc.; i.e., the same
for all the sites $i$. And at half filling, by symmetry, w$_{1} =
\frac{1}{2}$, so that $\Delta$ w$_{i}=0$. Hence $\Sigma_{n}$ becomes a
single explicit {\it function} of $G_n$ given by
\begin{equation}
\Sigma_{n} = \frac{U}{2}-\frac{1}{2G_{n}}\left(  1 - \sqrt{1 +
(UG_{n})^{2} } \right) \mbox{.} \label{sguni}
\end{equation}
The DMFT self consistency condition eq. (\ref{latdyson}) is easily
solved in this limit simply by fourier transforming. The lattice
Green's function is
\begin{equation}
G_{\bf k}(i\omega_{n})=(i\omega_{n} +\mu -\Sigma_{n}-\epsilon_{\bf
k})^{-1} \label{gkuni}\;\mbox{.}
\end{equation}
Hence the local Green's function is
\begin{eqnarray}
G_{n} &\equiv& G_{0}(i\omega_{n} +\mu -\Sigma_{n})\nonumber \\
 &=&\int d{\epsilon}\;\frac{D_0(\epsilon)}{i\omega_{n} +\mu
-\Sigma_{n}-\epsilon} \hspace{.1cm} \mbox{,} \label{lguni}
\end{eqnarray}
where $D_0(\epsilon)$ is the  bare density of states(DOS) of the
band structure determined by $t_{ij}$. Most of the results reported
in this chapter have been obtained using the semi-circular DOS (SDOS)
characteristic of the Bethe lattice in infinite dimensions, for
which
\begin{equation}
D_0(\epsilon)=\frac{1}{2\pi
t^2}\sqrt{4t^2-\epsilon^2}\hspace{.1cm}\mbox{,}
\hspace{.2cm}|{\epsilon}|<2t \; \mbox{,}
\end{equation}
whence $G_0 (z)$, the complex Hilbert transform of the bare DOS,
can be obtained analytically, as
\begin{equation}
G_{0}(z_n)=\frac{2}{z_n+\sqrt{z_n^2-4t^2}}\mbox{,}
\label{sdoscht}
\end{equation}
with $z_n=i\omega_n+\mu -\Sigma_n$. We also report some results
using the actual density of states for a square lattice whence
$G_0(z)$ has to be evaluated numerically. We will refer to these as
2DDOS results. In either case, the local Green's function and the
self energy  can be obtained by self consistently solving
eqns.~(\ref{sguni}) and (\ref{lguni}).

In the uniform case at half filling the temperature completely drops
out of the problem, as the $\ell$ occupancy probabilities are fixed
by symmetry. The metal-insulator transition we alluded to above
\cite{Dong1} hence takes place at a fixed value of $U$ independent
of temperature within the DMFT. For the SDOS case this value is $U =
2t$. The temperature dependence of the $b$ electron DOS can be
restored by including spatial fluctuations using approximations that
go beyond DMFT \cite{HRK_1}.

As the temperature is lowered, the homogeneous (disordered) phase
becomes unstable with respect to an ordered phase where both the $b$
and $\ell$ charge densities with spatial modulations at some
ordering wave-vector $\bf q$ develop; i.e.,  $\sum_{i}exp{(i\bf q
\cdot {\bf R}_{i}})<b_{i}^{\dag}b_{i}>$ and $\sum_{i}exp{(i\bf q
\cdot {\bf R}_{i}})<\ell_{i}^{\dag}\ell_{i}>$ acquire nonzero
values. If the transition to this ordered phase is continuous
(second-order), then it occurs when the charge-susceptibility (at
the ordering wave vector) diverges. The static, or zero (external)
frequency $b$-charge-susceptibility at the ordering wave vector
${\bf q}$ is obtainable as the Fourier transform of the
$b$-charge-density - $b$-charge-density correlation function:

\begin{eqnarray}
\chi({\bf q},T)& \equiv & -\frac{1}{N}\sum_{j,k}
e^{i{\bf q} \cdot({\bf R}_{j}-{\bf R}_{k})} \times \nonumber \\
&& \int_{0}^{\beta}d{\tau}
< b_{j}^{\dag}(\tau) b_{j}(\tau) b_{k}^{\dag}(0)b_{k}(0)>\, , \\
\nonumber & \equiv & T \sum_{n=-\infty}^{\infty} {\bar \chi}_n
({\bf q})\, , \label{suscep1}
\end{eqnarray}
Here $n$ indexes the Matsubara frequency of the fermionic
propagators that arise in a diagrammatic perturbation theory
evaluation of $\chi$. Using the generalized Dyson's
equation\cite{Brandt_2} valid for two particle propagators, one can
relate ${\bar \chi}_{n}$ to the simplest ``bubble graph
susceptibility'' ${\bar \chi}_{n}^{0}$ in the form
\begin{equation}
{\bar \chi}_{n}({\bf q}) = {\bar \chi}_{n}^{0}({\bf q})-{\bar
\chi}_{n}^{0}({\bf q}) \sum_{m=-\infty}^{\infty} \frac{\partial
\Sigma_{n}\left[ G \right]}{\partial G_{m}} {\bar \chi}_{m}({\bf
q}) \, ,
\end{equation}
where ${\bar \chi}_{n}^{0}$ is given by\cite{Brandt_2}
\begin{eqnarray}
{{\bar \chi}_{n}}^{0}({\bf q})&=&-\sum_{\bf k} G_{n}({\bf k} +{\bf
q})G_{n}({\bf k}) \,, \\
\nonumber &=& - \frac{1}{\sqrt{1-X^2}} \int_{-\infty}^{\infty}
d{\epsilon} \frac{D_0(\epsilon)} {i\omega_n  + \mu-\Sigma_n -
\epsilon}\times \nonumber \\
& &G_0 \bigg[ \frac{i\omega_n + \mu - \Sigma_n - X
\epsilon}{\sqrt{1-X^2}} \bigg]\, .
\end{eqnarray}
Here $D_0({\epsilon})$ is the bare DOS as before, and all of the
wave-vector dependence occurs via the quantity
\begin{equation}
X({\bf q}) \equiv  \frac{1}{d} \sum_{j=1}^{d}cos (q_{j})\, .
\end{equation}
and we use
\begin{equation}
G_0(z) \equiv \int d\epsilon D_0(\epsilon)/(z-\epsilon)
\label{doscht}
\end{equation}
to denote the complex Hilbert Transform of the bare DOS as before.
Similarly, the $\ell$-$\ell$ charge correlation function defined by
\begin{equation}
{\chi}_{{\ell}-{\ell}}({\bf q}, T)\equiv
\frac{1}{NT}\sum_{j,k}e^{i{\bf q}\cdot({\bf R}_{j}-{\bf R}_{k})} <
n_{\ell j}n_{\ell k}>
\end{equation}
can be shown\cite{Brandt_2} to diverge at the same temperature [for
the same value of $X({\bf q})$] as the $b$-charge-susceptibility
discussed above. Therefore, a divergence of the
$b$-charge-susceptibility [$\chi({\bf q}, T)$ in eq.
(\ref{suscep1})] also signals an instability towards the generation
of long-range order in the $\ell$ charge density corresponding to
the same parameter $X$.

\begin{figure}[tbp]
\begin{center}
\includegraphics[scale=0.5]{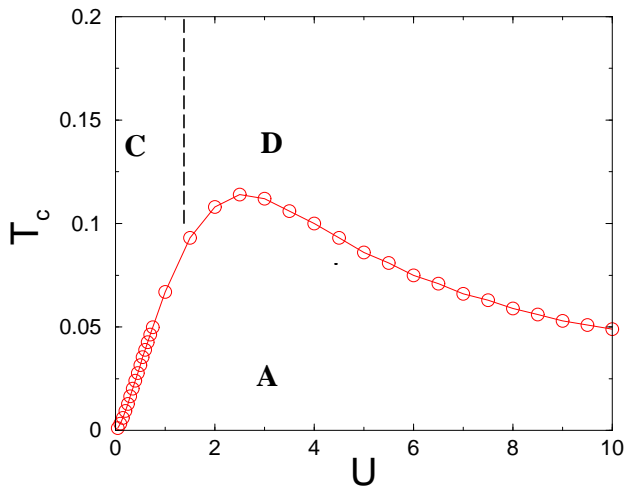}
\end{center}
\caption{(Color online)Conventional phase diagram of the half-filled SFKM in the
$T-U$ plane. Phase A, C, and D label the  CDW phase, the uniform
non-Fermi liquid metal, and the uniform insulator respectively.}
\label{ophase}
\end{figure}

The mapping  ${\bf q}\rightarrow X({\bf q})$ is a many-to-one
mapping that determines an equivalence class of wave vectors in the
Brillouin zone. In the $d \rightarrow \infty$ limit, ``generic''
wave vectors are all mapped to $X=0$, since $cos{q_j}$ can be
thought of as a random number between -1 and 1 for generic points in
the Brillouin zone. All the possible values of $X \, (-1\le X\le 1)
$ can be sampled, however, using a wave vector that lies on the
diagonal of the first Brillouin zone extending from the zone center
($X=1$) to the zone corner ($X=-1$). An ordering at the zone corner
$[X=-1,{\bf Q}=(\pi, \pi, ...)$] corresponds to a two sublattice
(checkerboard) state with the $\ell$ electrons preferentially
occupying one sublattice [$exp(i{\bf Q}.R_{j})=1$] and the $b$
electrons preferentially occupying the other sublattice [$exp(i{\bf
Q}.R_{j})=-1$]. Here we only consider the $X=-1$ case. The
susceptibility $\chi({\bf q}, T)$ then diverges at the temperature
 $T_{c}$  $(X=-1)$  and $T_c$ as a function of $U$ for the SDOS case is
shown in Fig.~\ref{ophase}. Phase $C$ denotes the non-Fermi liquid
metal ($\Im\Sigma(\omega=0)\ne 0$ and temperature independent), $D$
is the uniform insulator and $A$ is the two-sublattice or
checkerboard charge-density-wave(CDW) phase\cite{Footnote}.

\section{ Spectral and transport properties of the Charge-density-wave phase}
%\noindent

The checkerboard charge-density-wave(CDW) phase has two inequivalent
sublattices, $A$ and $B$, defined by:
\begin{eqnarray}
exp(i{\bf Q} \cdot R_{i})=+1 \,\,\, \mbox{ for i belonging to A}\\
\nonumber exp(i{\bf Q} \cdot R_{i})=-1 \, \, \,\mbox{ for i
belonging to B}
\end{eqnarray}
with ${\bf Q}=(\pi,\pi,\pi,.....)$. The two sublattices
spontaneously develop different electronic densities relative to
each other, but all sites of a given sublattice are equivalent. All
the quantities such as w$_{1i}$, $\Sigma_{n}^{ii}$, $G_{n}^{ii}$,
etc., introduced earlier in the DMFT formalism section now become
double valued with respect to $i$. Hence we can write
\begin{equation}
\mbox{w}_{1i} = \left \{ \begin{array}{ll} \mbox{w}_{1a} &
\mbox{if $i$ belongs to sublattice  $A$ }\\
\mbox{w}_{1b} & \mbox{if $i$ belongs to sublattice $B$ }
\end{array} \right. \mbox{.}
\end{equation}
Similarly, for the local self-energy and Green's function, one has
\begin{equation}
\Sigma_{n}^{ii} = \left \{ \begin{array}{ll} \Sigma_{n}^{A}&
\mbox{if $i$  belongs to sublattice  $A$ }\\
\Sigma_{n}^{B} & \mbox{if $i$ belongs to sublattice  $B$}
\end{array} \right. \mbox{;}
\end{equation}
\begin{equation}
G_{n}^{ii} = \left \{ \begin{array}{ll} G_{n}^{A}&
\mbox{if $i$ belongs to sublattice  $A$ }\\
G_{n}^{B}& \mbox{if $i$ belongs to sublattice  $B$}
\end{array} \right. \mbox{.}
\end{equation}
At half filling, by symmetry,
\begin{equation}
(\mbox{w}_{1a}- \frac{1}{2})= -(\mbox{w}_{1b}-\frac{1}{2}) \equiv\Delta \mbox{w}
\mbox{.}
\end{equation}
Hence, the relation eq. (\ref{Sigmaii}) becomes
\begin{equation}
\Sigma_{n}^{A,B}=\frac{U}{2}-\frac{1}{2G_{n}^{A,B}}\left(  1 -
\sqrt{1 + (UG_{n}^{A,B})^{2} \pm 4 (\Delta \mbox{w}) UG_{n}^{A,B}}\right)
\mbox{,} \label{sigab}
\end{equation}

To complete the DMFT self consistency loop, we have to solve the
lattice Dyson equation eq. (\ref{latdyson}) in the present, two
sublattice CDW context. For this purpose it is convenient to
introduce a two-component spinor field $\psi_{\bf k}$,
\begin{equation}
\psi_{\bf k} \equiv \left( \begin{array}{l} b_{{\bf k}A}\\
b_{{\bf k}B}\end{array}\right)\mbox{,}
\end{equation}
with
\[ b_{{\bf k}A} \equiv \sum_{i \epsilon A} b_i \exp(i {\bf k}.{\bf R}_{i})~, \]
etc., where ${\bf k}$ now only takes values within the reduced
Brillouin zone corresponding to the A sublattice alone. In terms of
the matrix Green's function defined in the usual way as
\begin{equation}
{\bf G}_{{\bf k}n} \equiv  << \psi_{\bf k} ; \psi_{\bf k}^{\dag}
>>\mbox{,}
\end{equation}
the lattice Dyson equation eq. (\ref{latdyson}) reduces in the
present context to the simple 2 x 2 matrix equation
\begin{equation}
[(i\omega_{n} + \mu){\bf 1} - {\bf M}_{{\bf k}n}]{\bf G}_{{\bf k}n}
= {\bf 1}\mbox{,}
\end{equation}
where
\begin{equation}
{\bf M}_{{\bf k}n} = \left(\begin{array}{ll} {\Sigma_{n}^{A}}  &  \epsilon_{\bf k}\\
\epsilon_{\bf k} & {\Sigma_{n}^{B}} \end{array}\right)\mbox{.}
\end{equation}
Hence ${\bf G}_{{\bf k}n}$ is given by
\begin{equation}
{\bf G}_{{\bf k}n} = \left( \begin{array}{ll}{\xi_{n}^{A}} &
-\epsilon_{\bf k}\\-\epsilon_{\bf k} &
{\xi_{n}^{B}}\end{array}\right)^{-1}\mbox{,} \label{latticeG}
\end{equation}
with
\[\xi_{n}^{A,B} \equiv i\omega_{n}+ \mu  - \Sigma_{n}^{A,B}~.\]
The local Green's functions $G_{n}^{A,B}$ are just the diagonal
components of the matrix:
\begin{equation}
{\bf G}_{n} = \sum_{\bf k} {\bf G}_{{\bf k}n}\mbox{.}
\end{equation}
Hence we get, finally,
\begin{equation}
G_{n}^{A,B}=\xi_{n}^{B,A}\int \frac {D_0(\epsilon) d\epsilon }
{\xi_{n}^{A} \xi_{n}^{B} - {\epsilon}^{2}} =
\frac{G_{0}(z_{n})}{z_{n}} \xi_{n}^{B,A}\mbox{,}\label{GAB}
\end{equation}
where $D_0(\epsilon)$ is bare DOS of the $b$ band  as before,
\[ z_{n} \equiv \sqrt{\xi_{n}^{A} \xi_{n}^{B}}~, \]
and $ G_{0}(z)$ is the complex Hilbert Transform of the bare DOS as
defined before in eq. (\ref{doscht}). Again, for the case of the
semi-circular DOS the calculations become simpler as $G_{0}(z)$ is
analytically known (cf eq. (\ref{sdoscht})).

We note that $\Delta$w can be thought of as the order parameter
characterizing the CDW phase. In the uniform phase, $\Delta$w$= 0$,
and $\Sigma_{n}^{A} = \Sigma_{n}^{B}$, etc., and all the above
equations reduce to the corresponding equations discussed earlier
for the uniform case.

\section{Results and Discussion}
%\noindent

\begin{figure}[tbp]
\begin{center}
\includegraphics[scale=0.4]{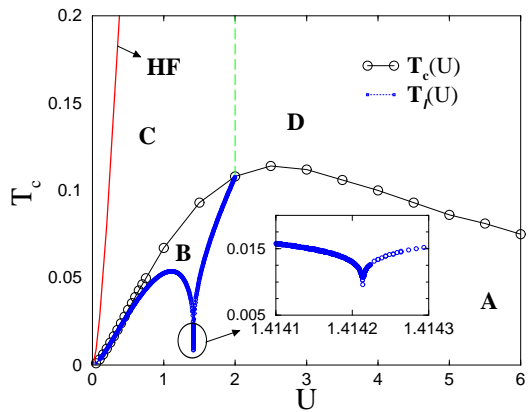}
\end{center}
\caption{(Color online) New phase diagram of the half-filled SFKM in the $T-U$
plane. Phase A, B, C, and D represent gapped CDW, gapless CDW,
uniform non-Fermi liquid metal, and uniform insulator respectively.}
\label{Newphase}
\end{figure}

\begin{figure}[tbp]
\begin{center}
\includegraphics[scale=0.5]{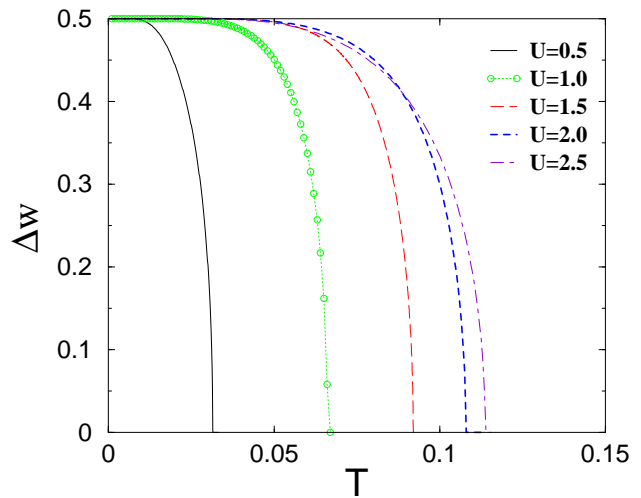}
\end{center}
\caption{(Color online) Variation of the order parameter ($\Delta$w) with
temperature for different values of $U$. (All the energies are
scaled in units of $t$.)} \label{chap2figb}
\end{figure}

Next we discuss in detail our results for the DMFT spectral and
transport properties in the CDW phase of the half-filled SFKM. For
convenience we measure all energies such as $\omega$, $U$, {\it
etc.} in units of $t$ for the SDOS, and $2t$ for the 2DDOS.
Consequently, the bare $b$ band width on the Bethe lattice or on the
square lattice is $W=4$.

The main result from our study is summarized in the {\it new `phase
diagram'} for the half-filled SFKM,  shown in Fig.~\ref{Newphase}.
As noted before in the introductory section, there is a region in
the phase diagram, marked B in Fig.~\ref{Newphase} and not
emphasized in the literature before, where the CDW phase order
parameter is non-zero, {\it but there is no gap in the spectrum}.
This is unlike the Hartree-Fock (HF) mean-field theory prediction,
where the CDW phase is always gapped. We will refer to this as the
``gapless CDW phase''. Furthermore, the `transition temperature'
$T_l(U)$ between this novel `phase' and the conventional gapped
CDW phase has an interesting {\it reentrant} structure as U
increases, as depicted in Fig.~\ref{Newphase}.

\begin{figure*}
\begin{center}
\includegraphics[scale=0.32]{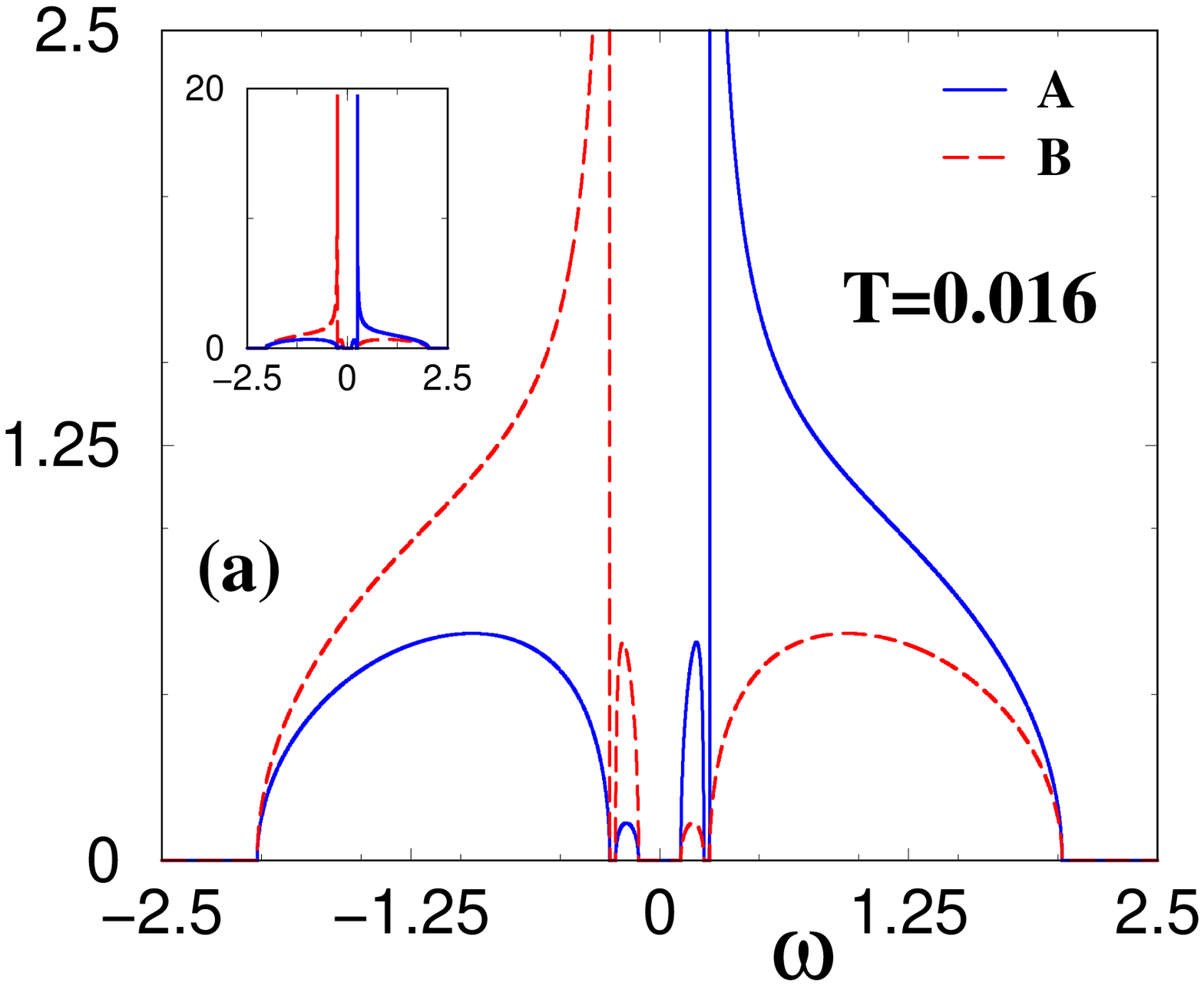}
\includegraphics[scale=0.32]{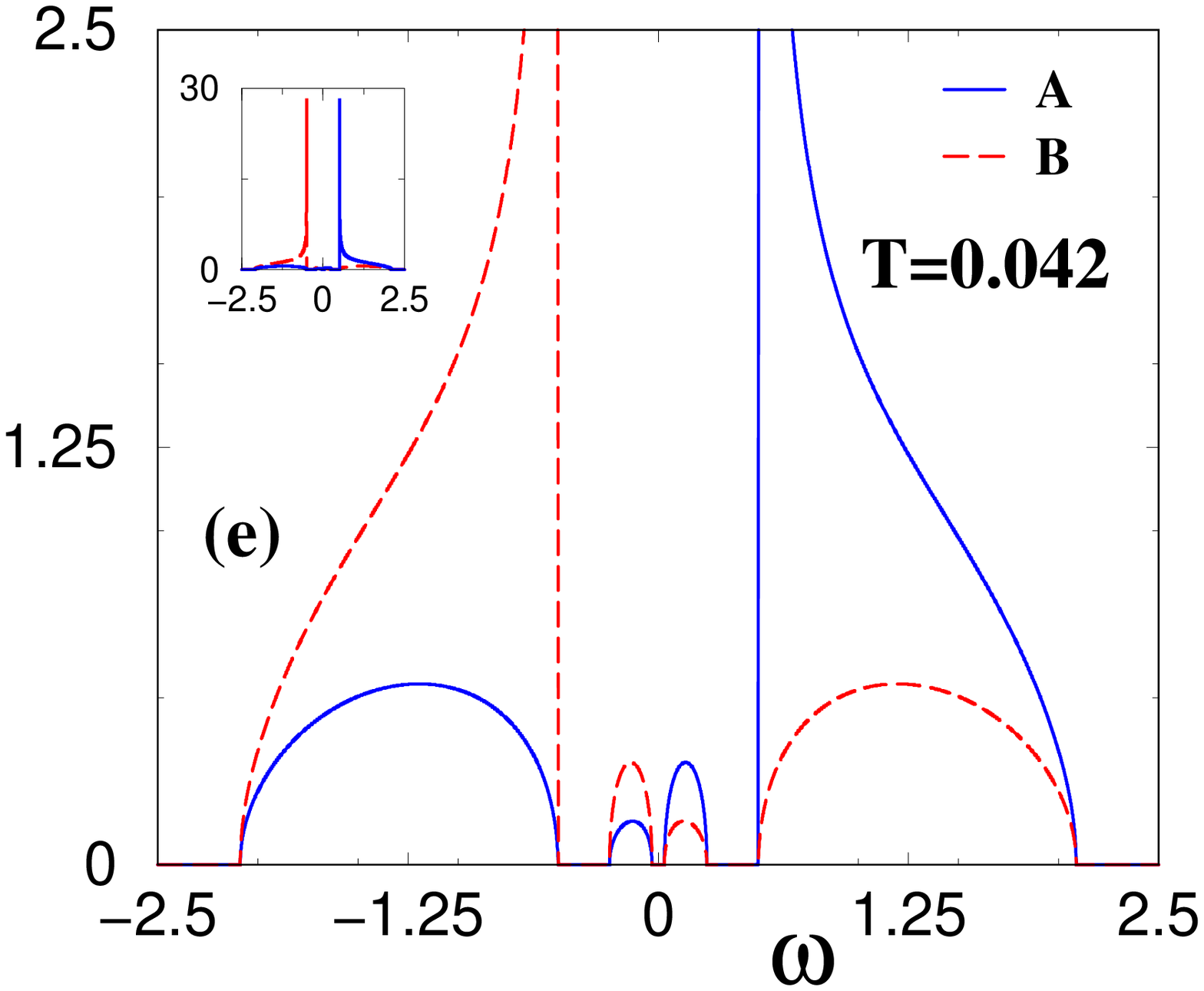}
\includegraphics[scale=0.32]{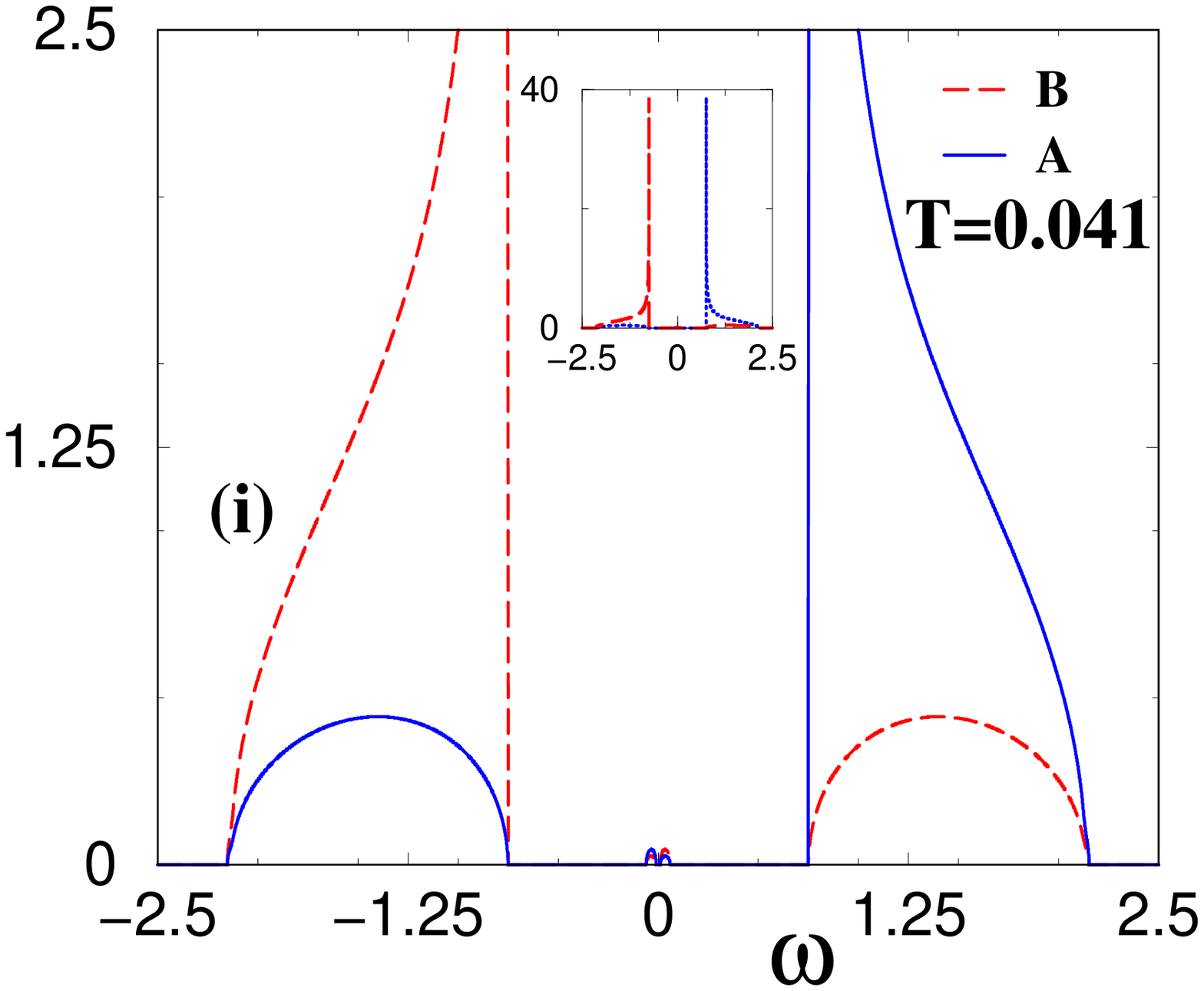}
\includegraphics[scale=0.32]{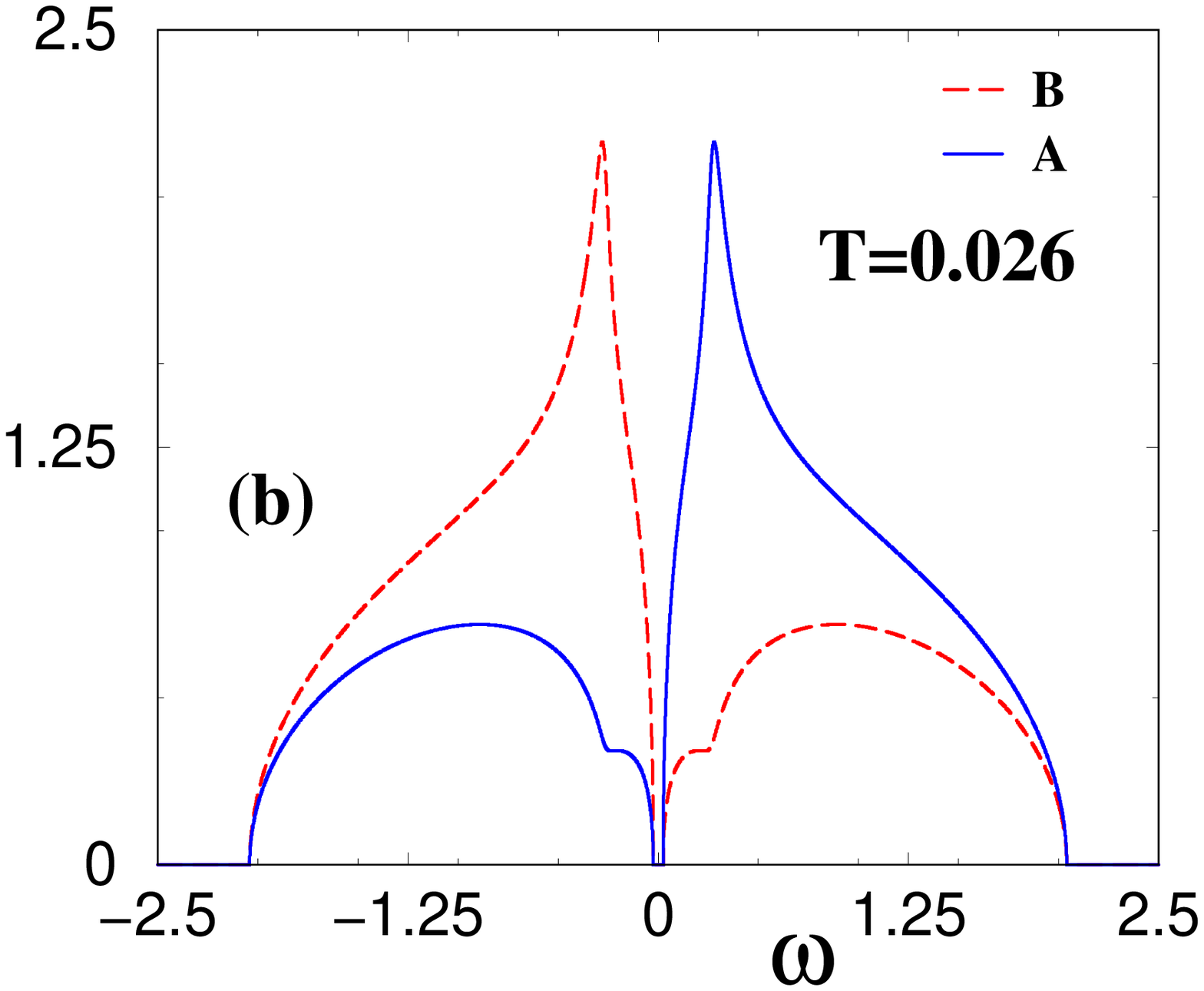}
\includegraphics[scale=0.32]{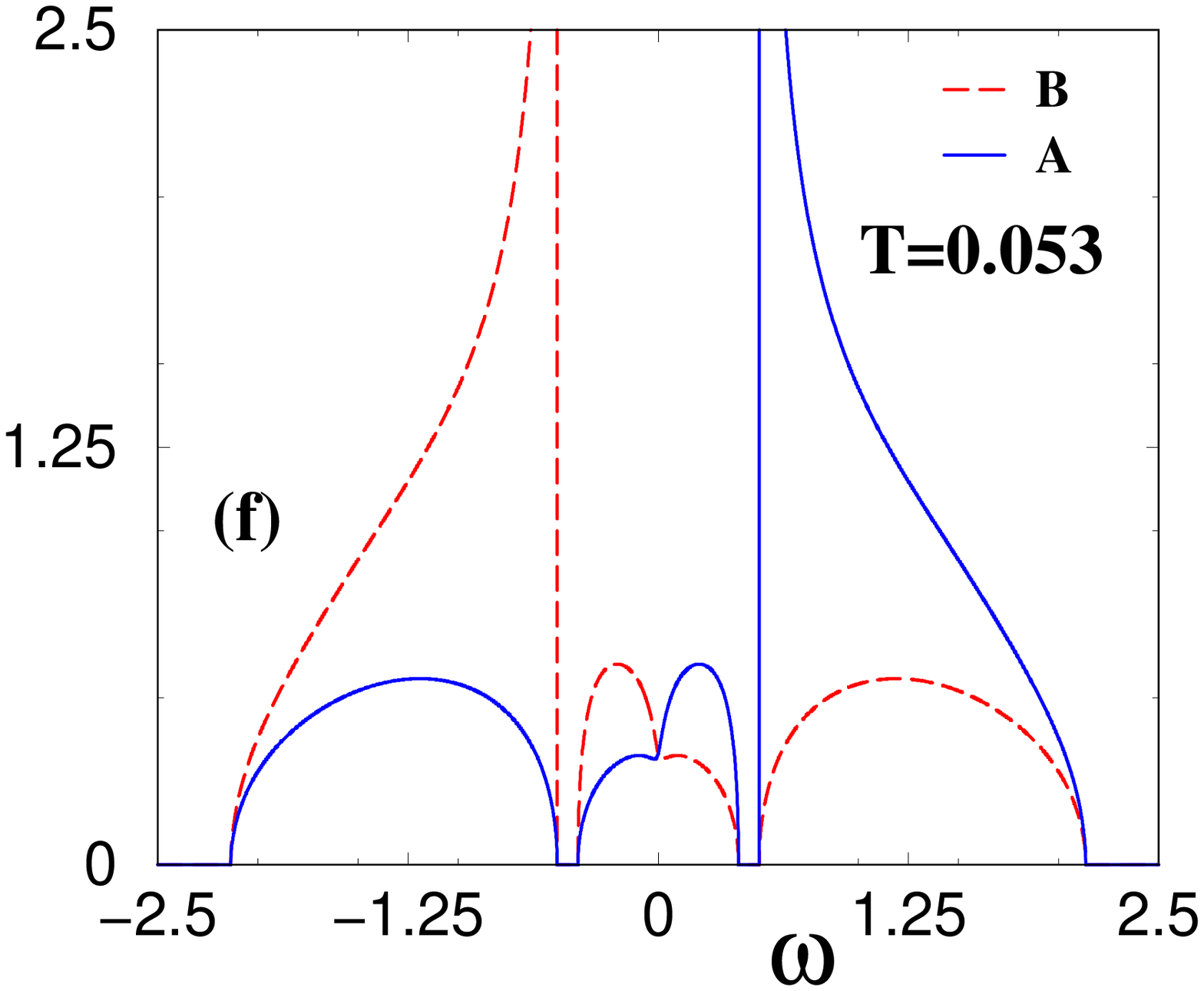}
\includegraphics[scale=0.32]{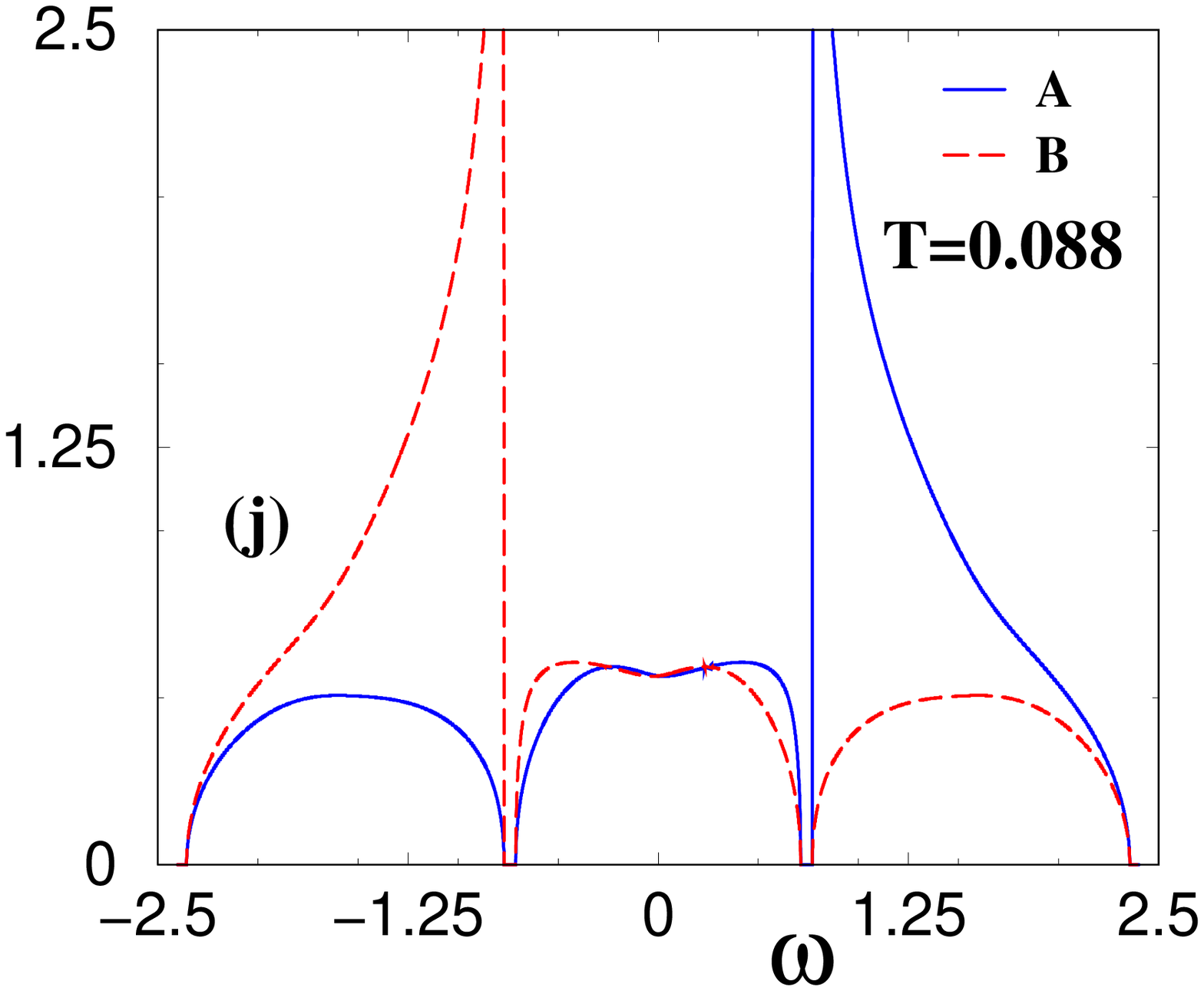}
\includegraphics[scale=0.32]{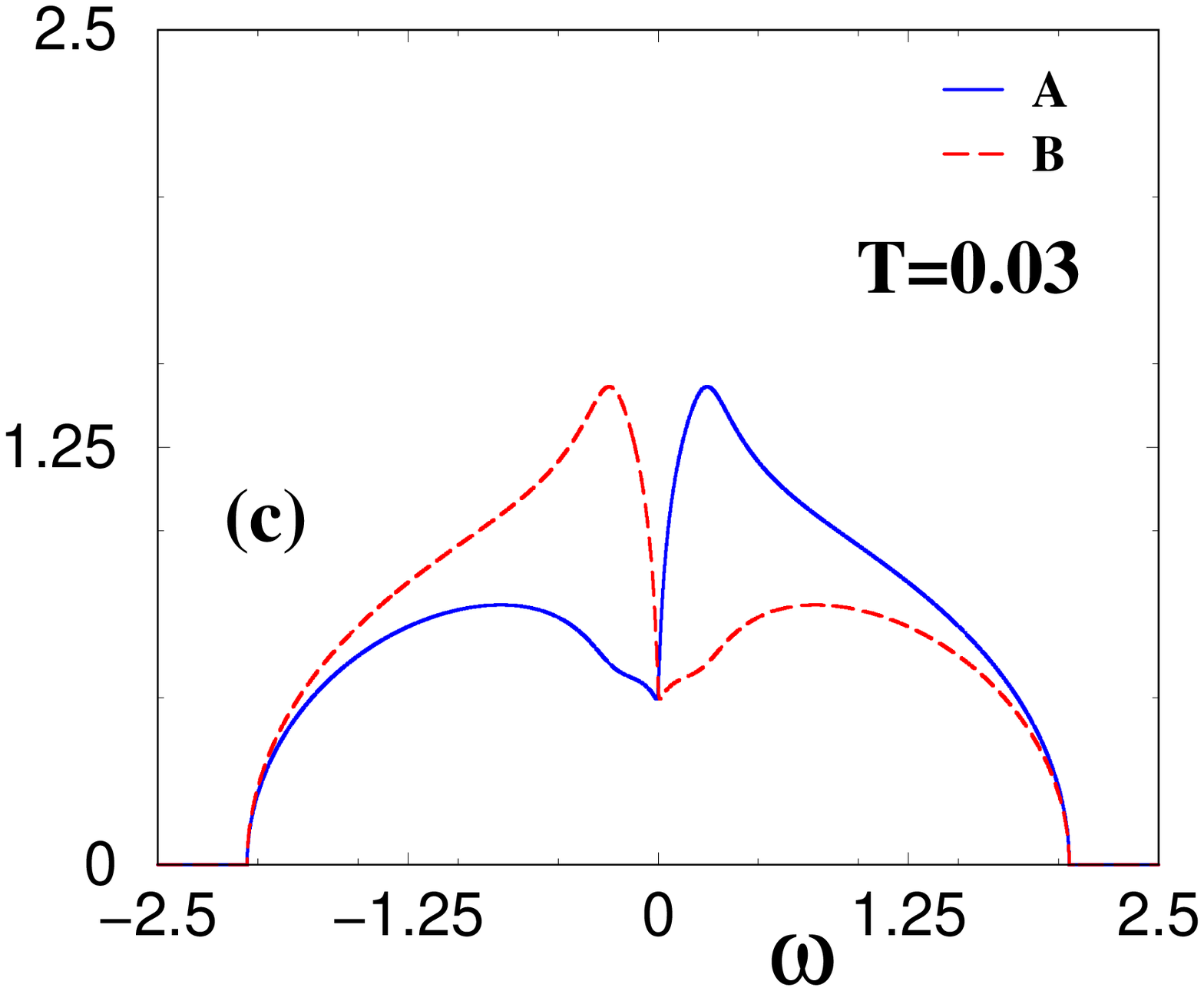}
\includegraphics[scale=0.32]{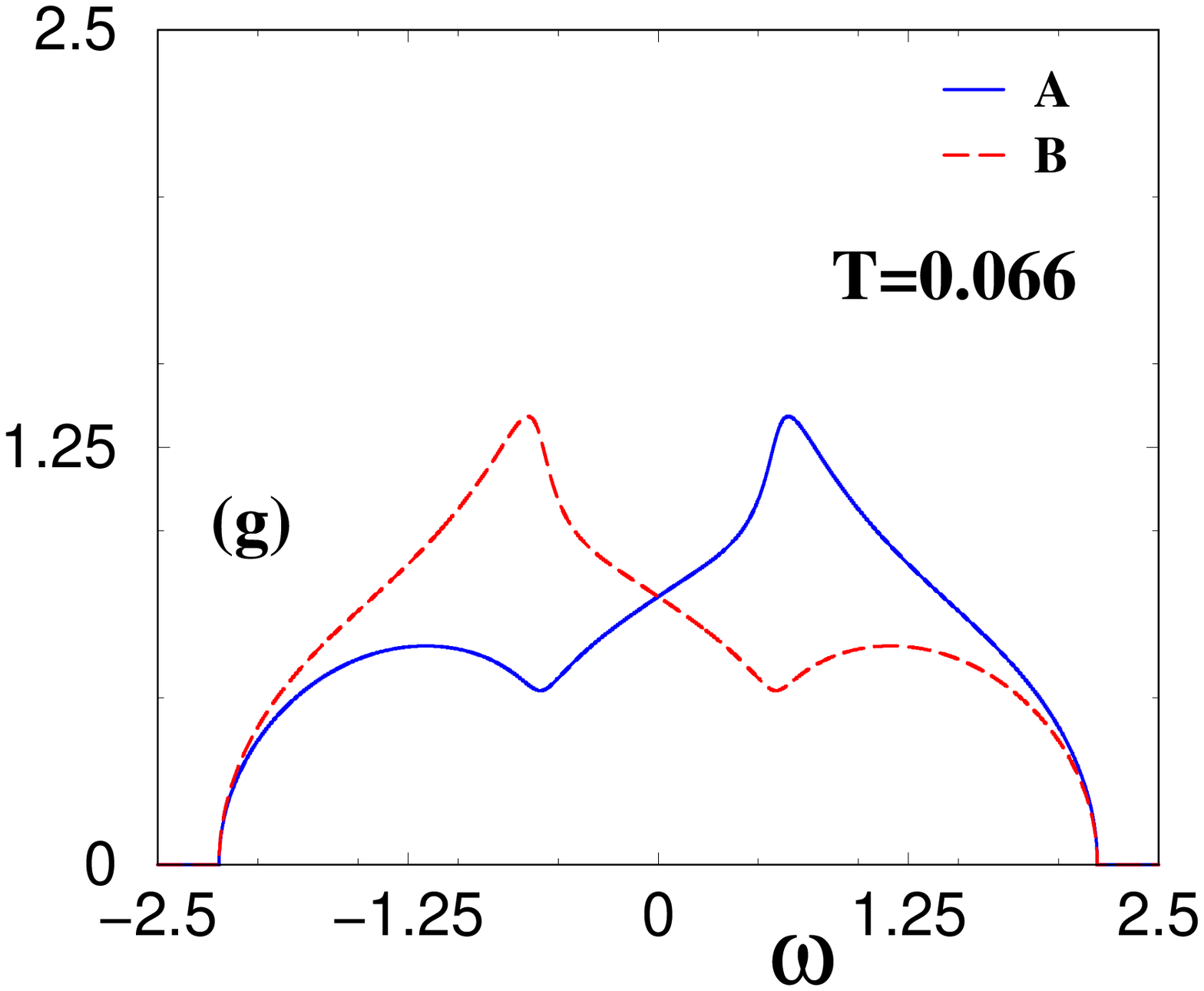}
\includegraphics[scale=0.32]{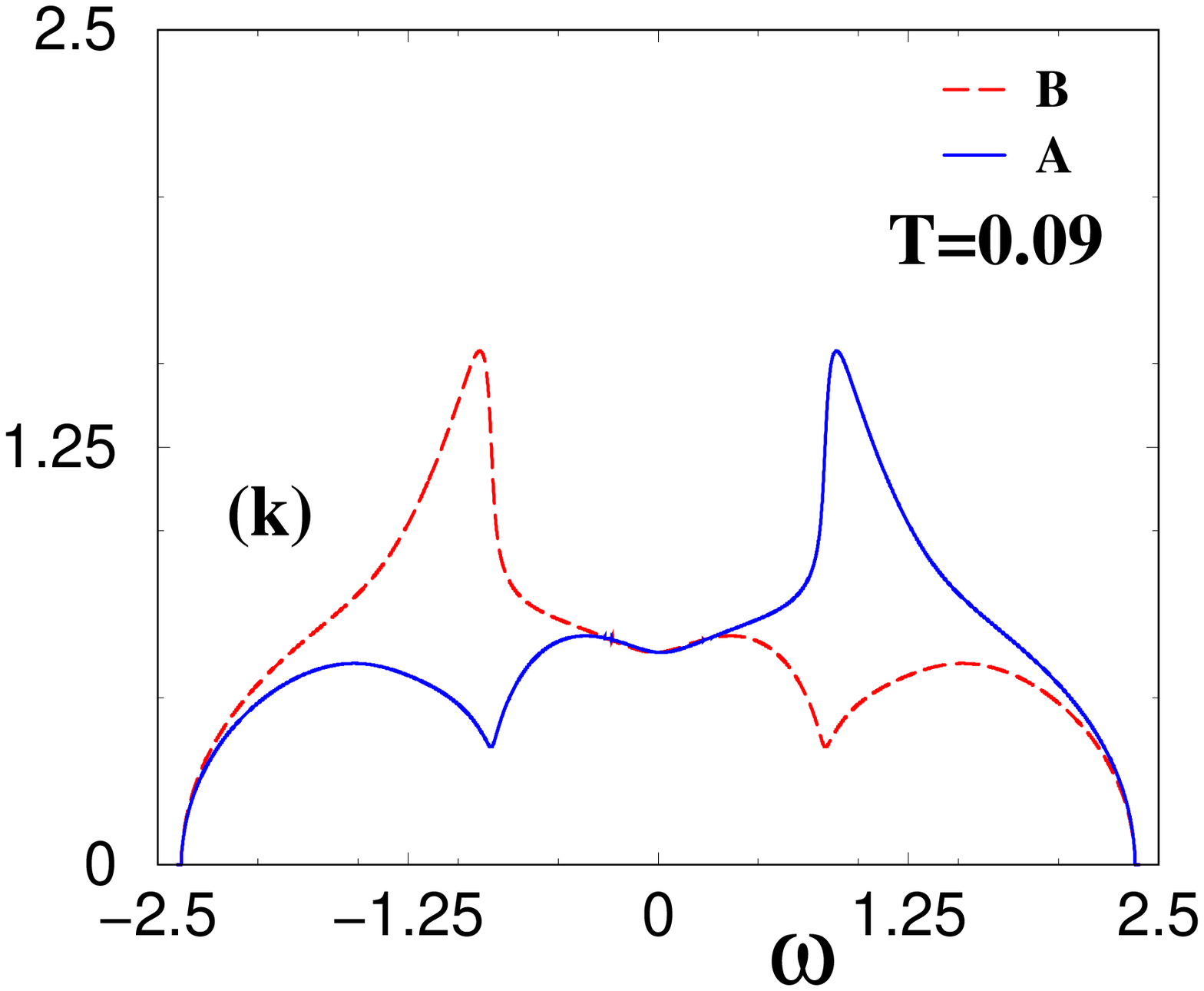}
\includegraphics[scale=0.32]{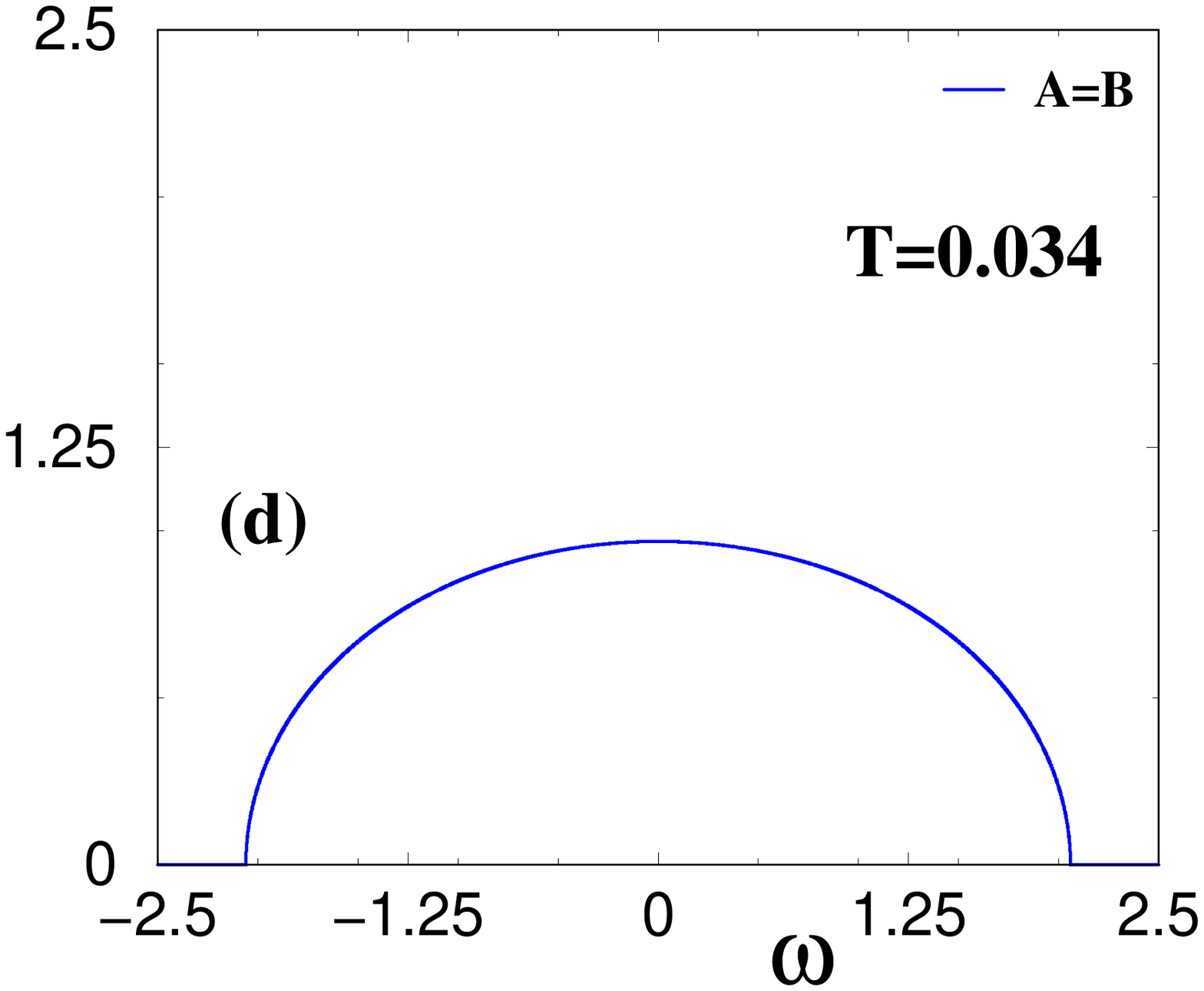}
\includegraphics[scale=0.32]{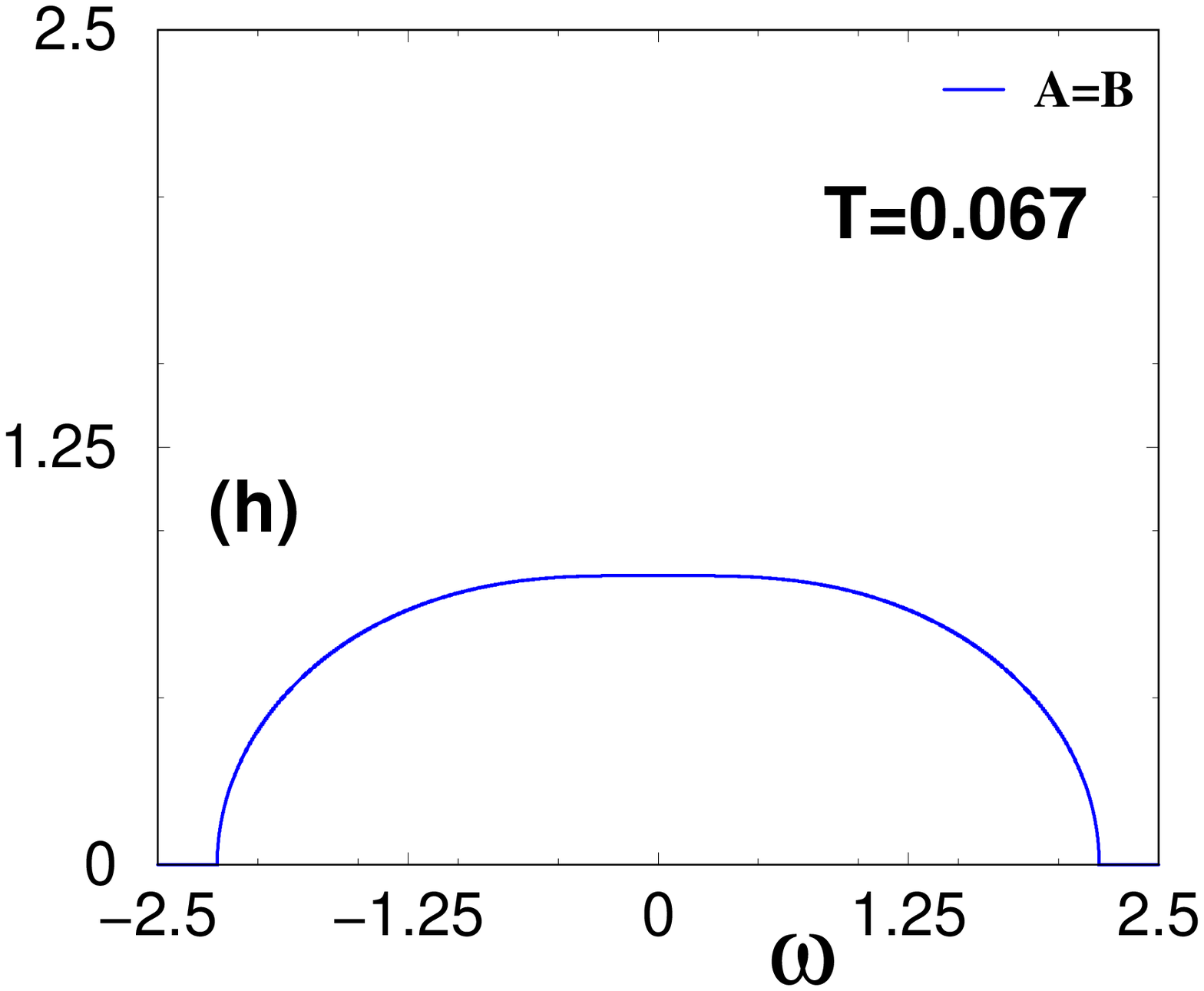}
\includegraphics[scale=0.32]{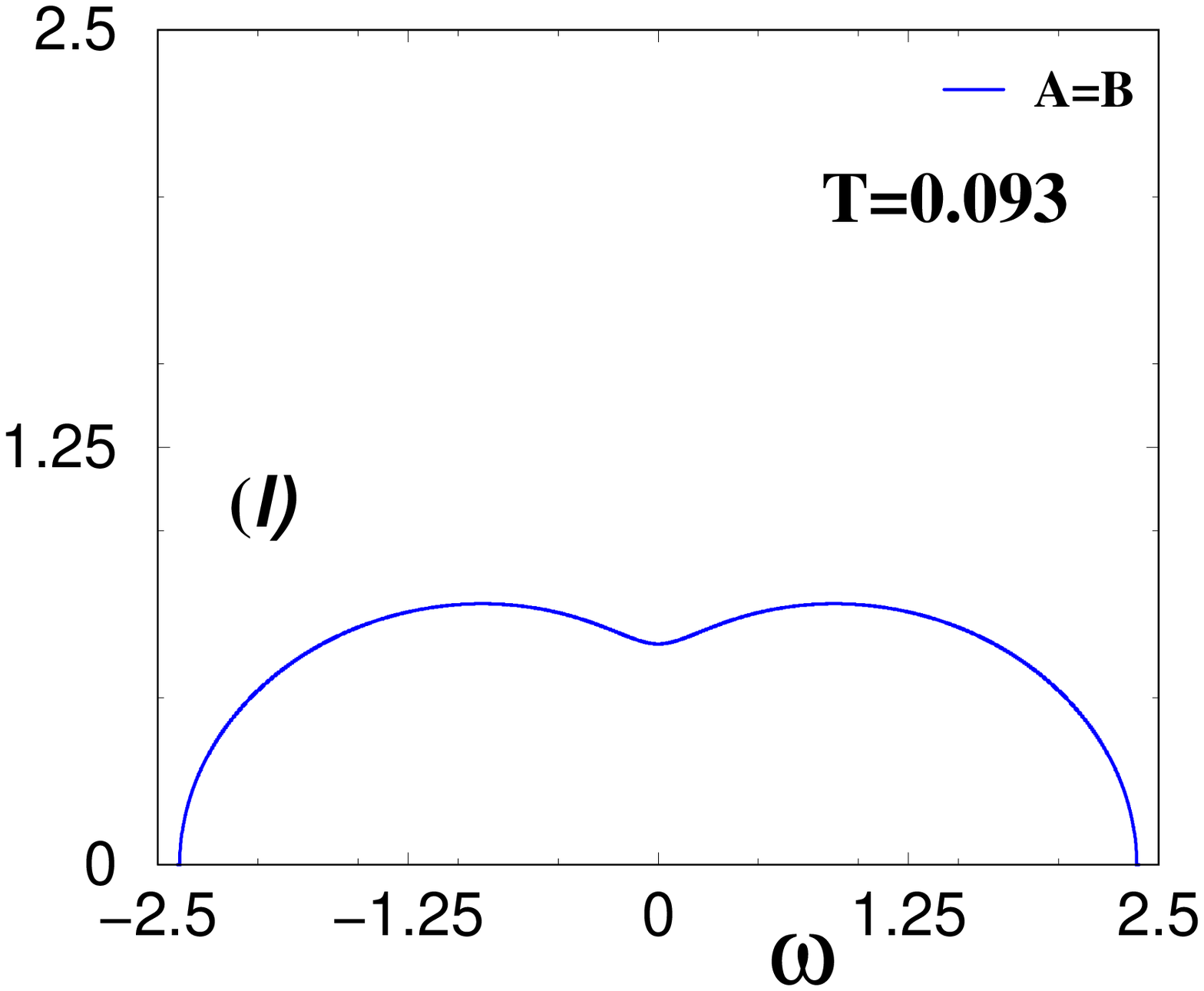}
\end{center}
\caption{(Color online) Diagrams showing the evolution of local spectral function
for SDOS as a function of temperature for three values of
$U=0.5,1.0,1.5$ in the first, second, and third columns
respectively, and in increasing order of temperature. Dotted lines
are for sublattice $A$ and long dashed lines are for sublattice
$B$.} \label{figsdos}
\end{figure*}

\begin{figure*}
\begin{center}
\includegraphics[scale=0.34]{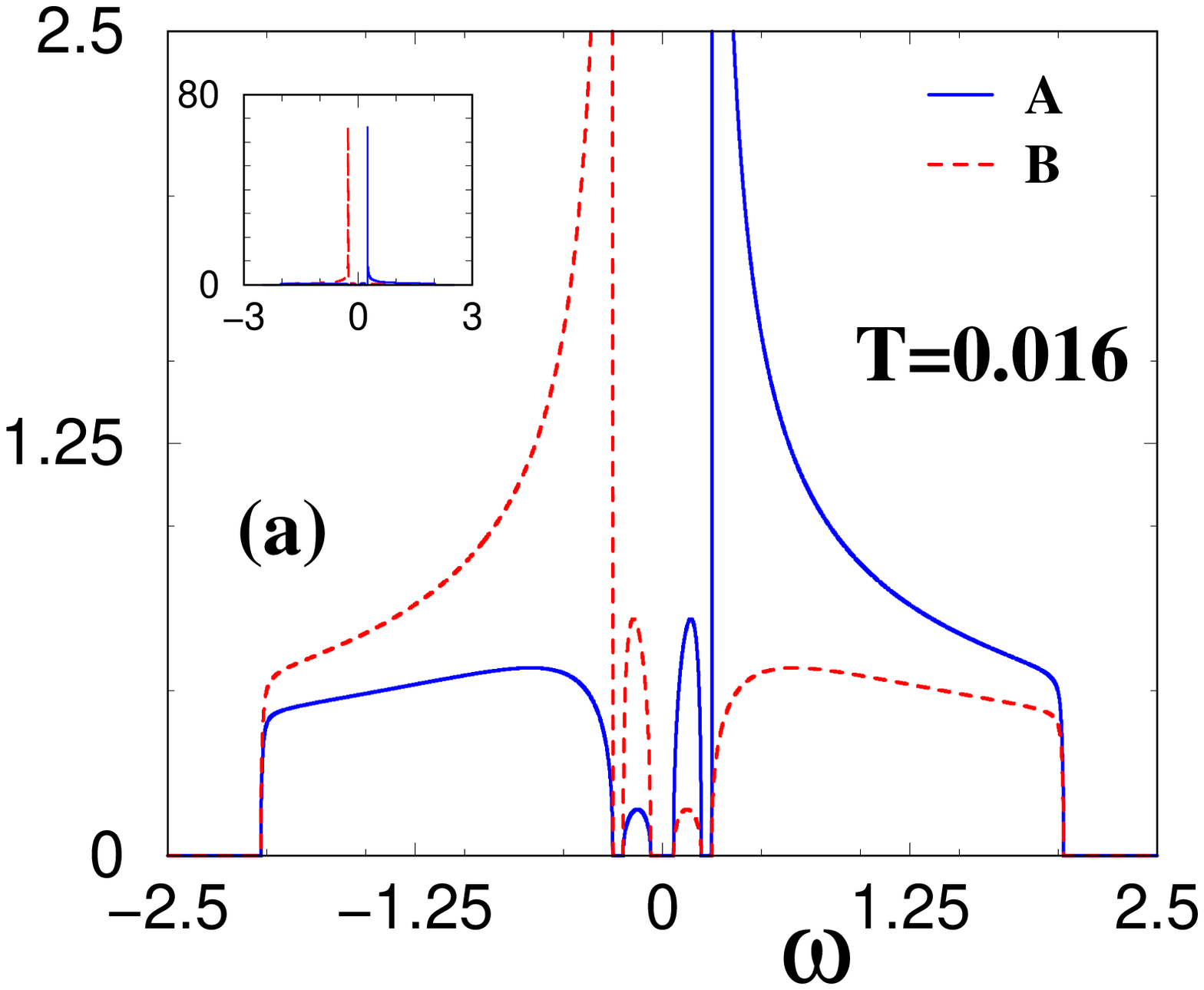}
\includegraphics[scale=0.34]{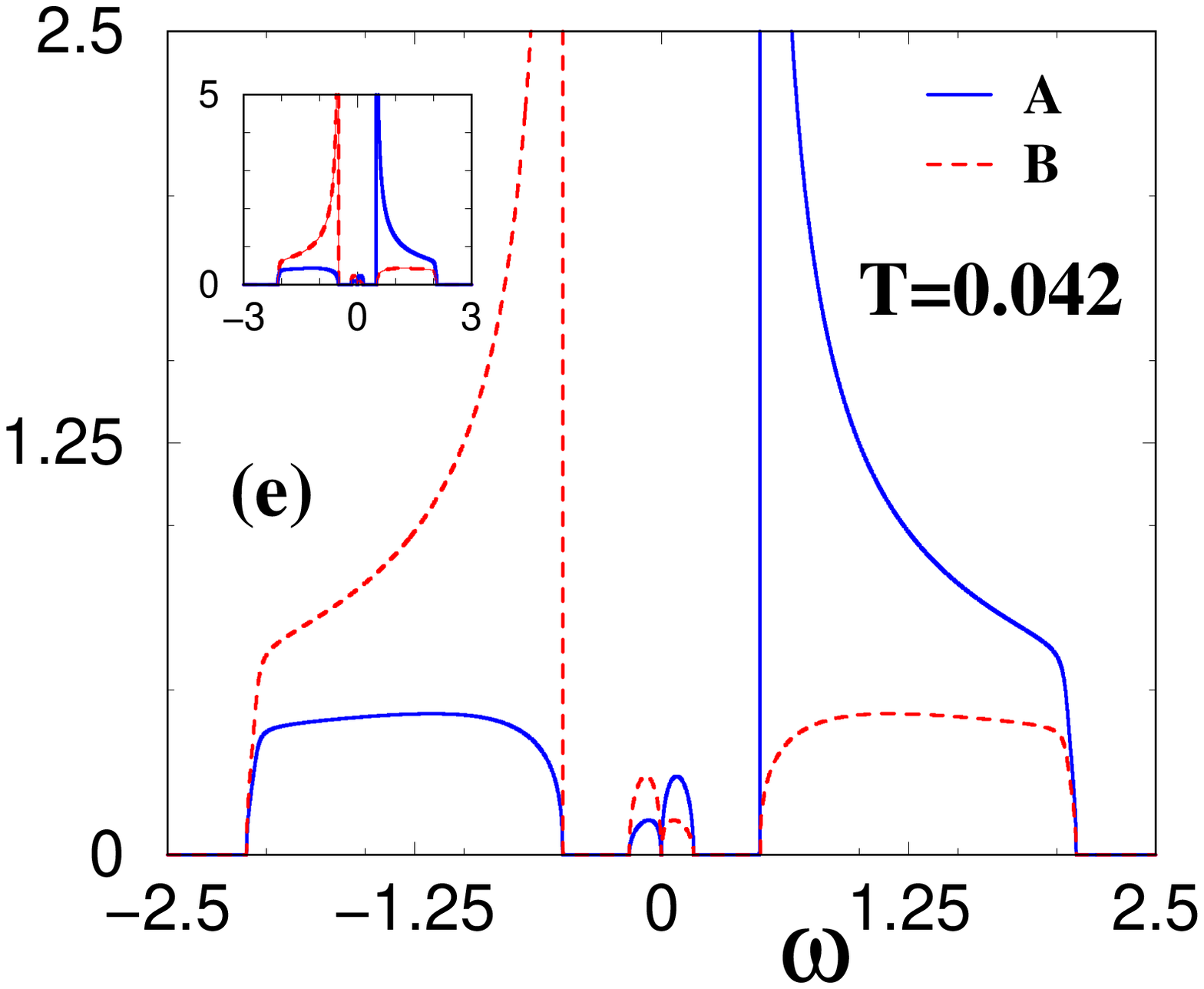}
\includegraphics[scale=0.34]{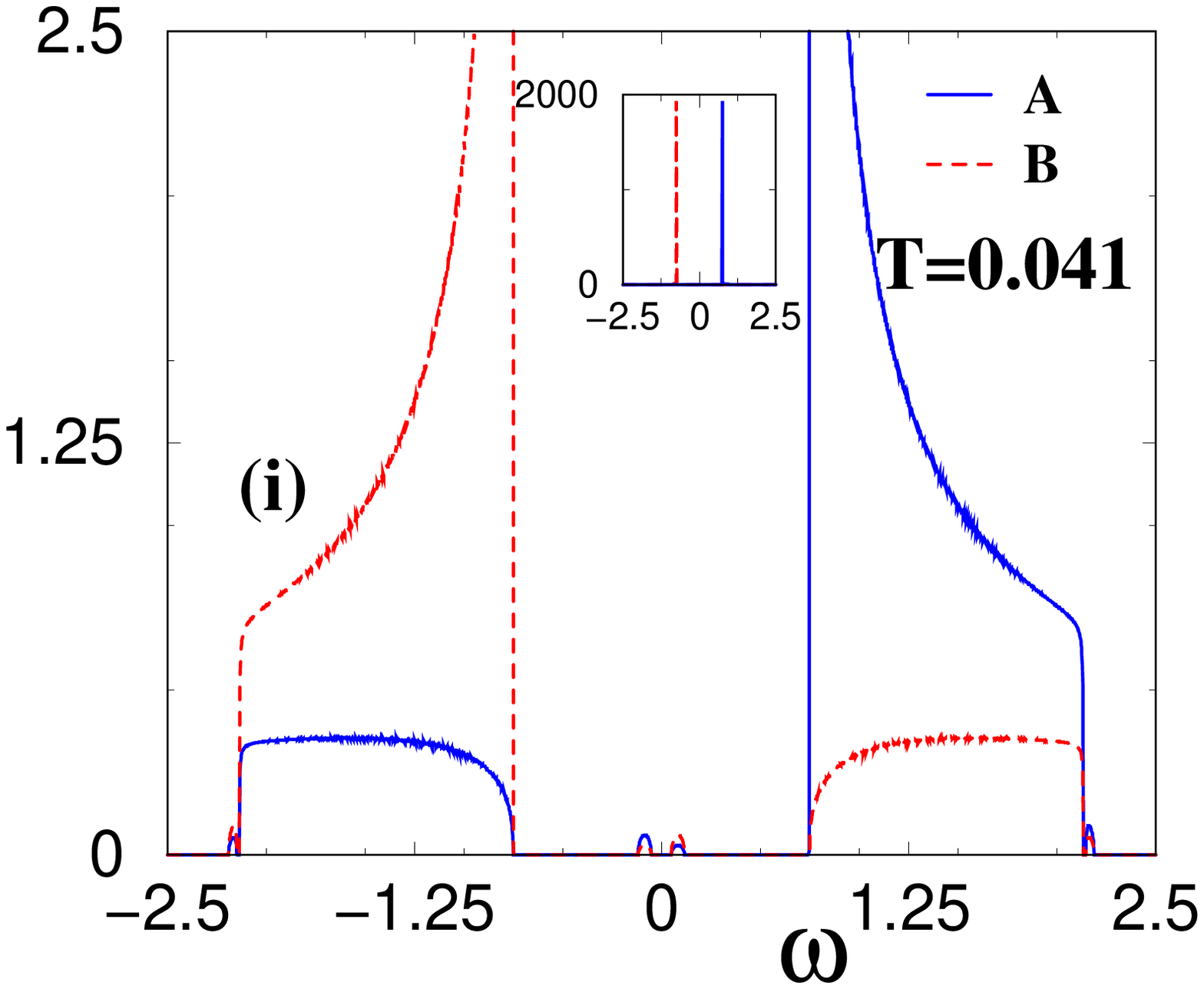}
\includegraphics[scale=0.34]{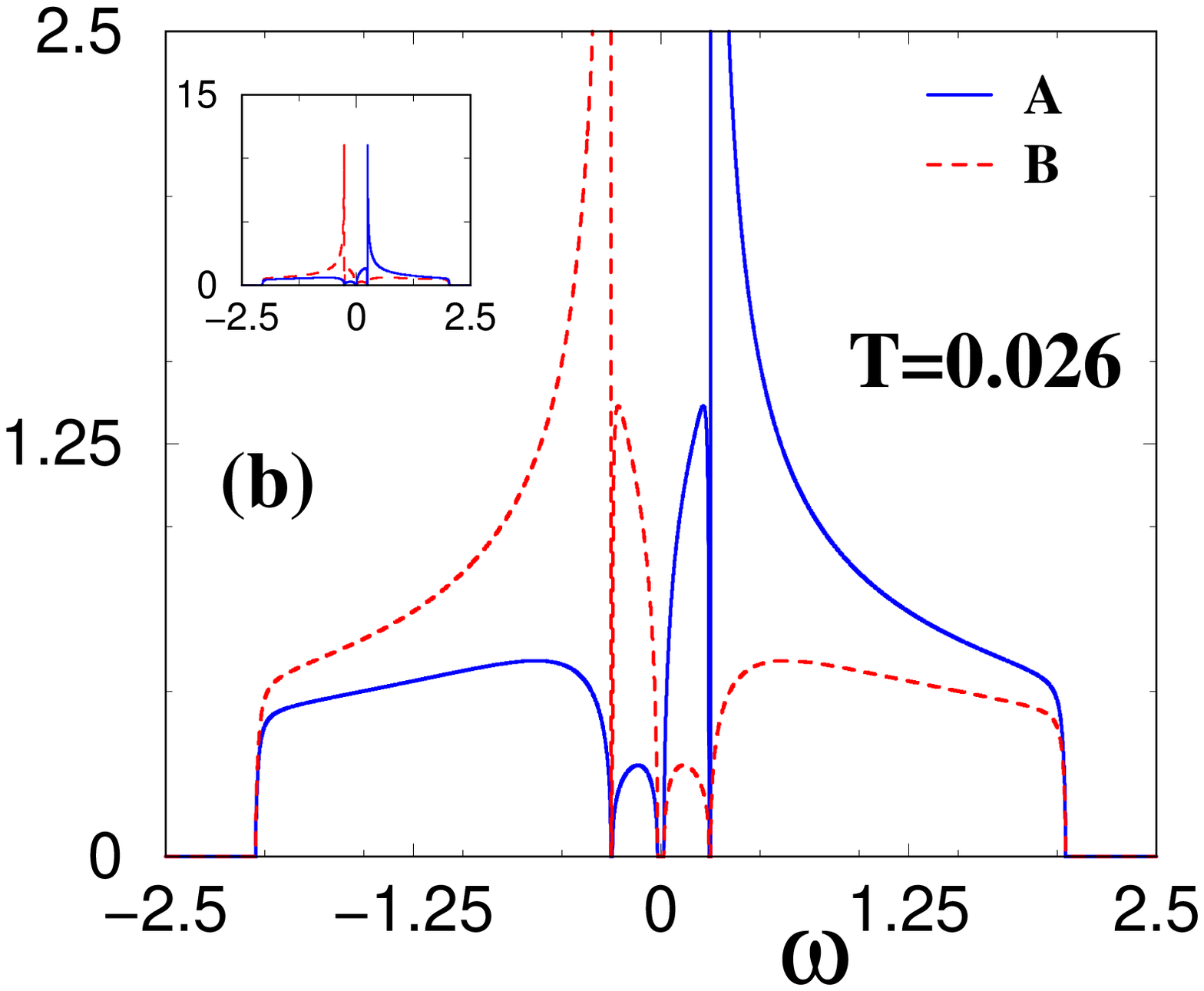}
\includegraphics[scale=0.34]{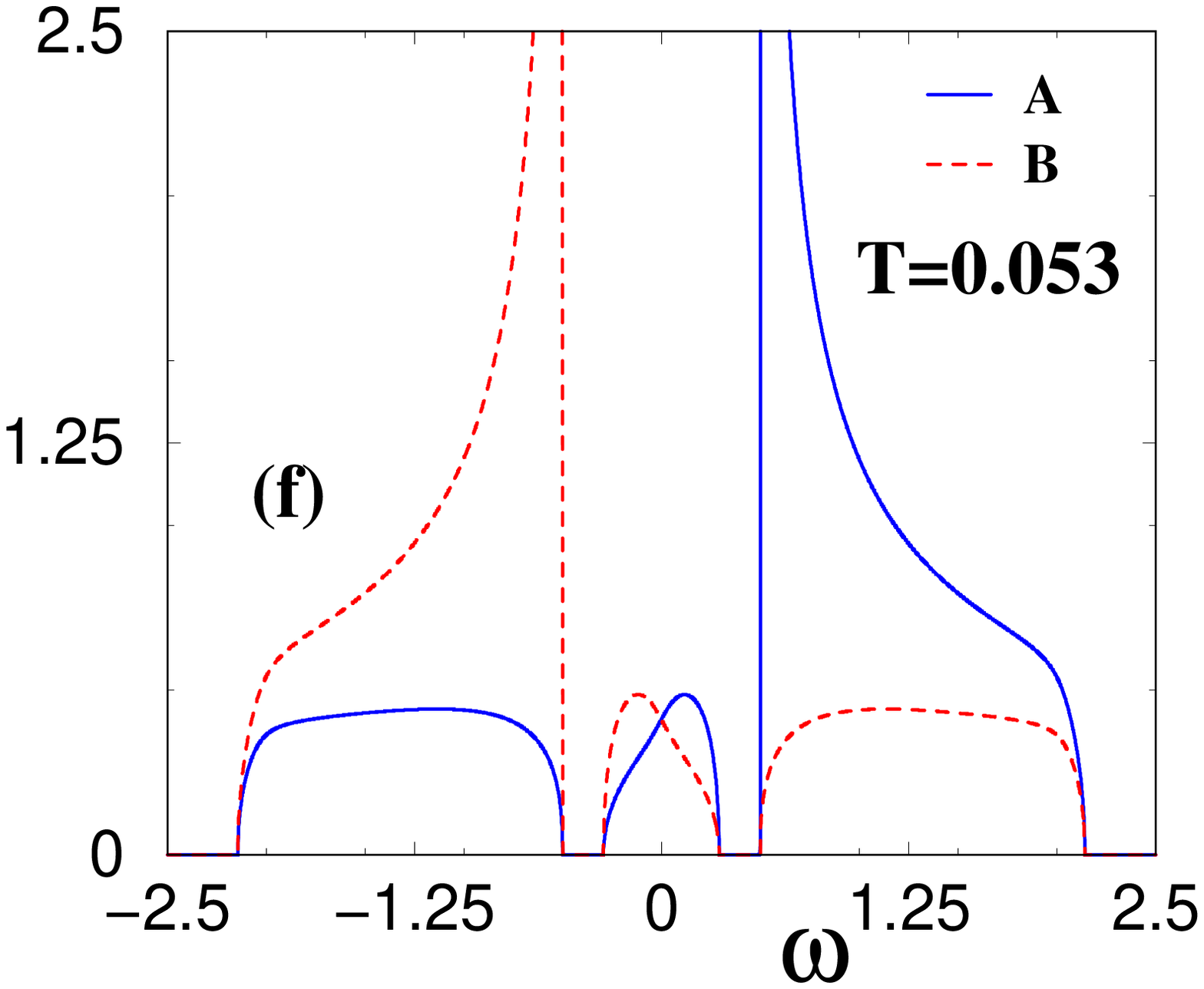}
\includegraphics[scale=0.34]{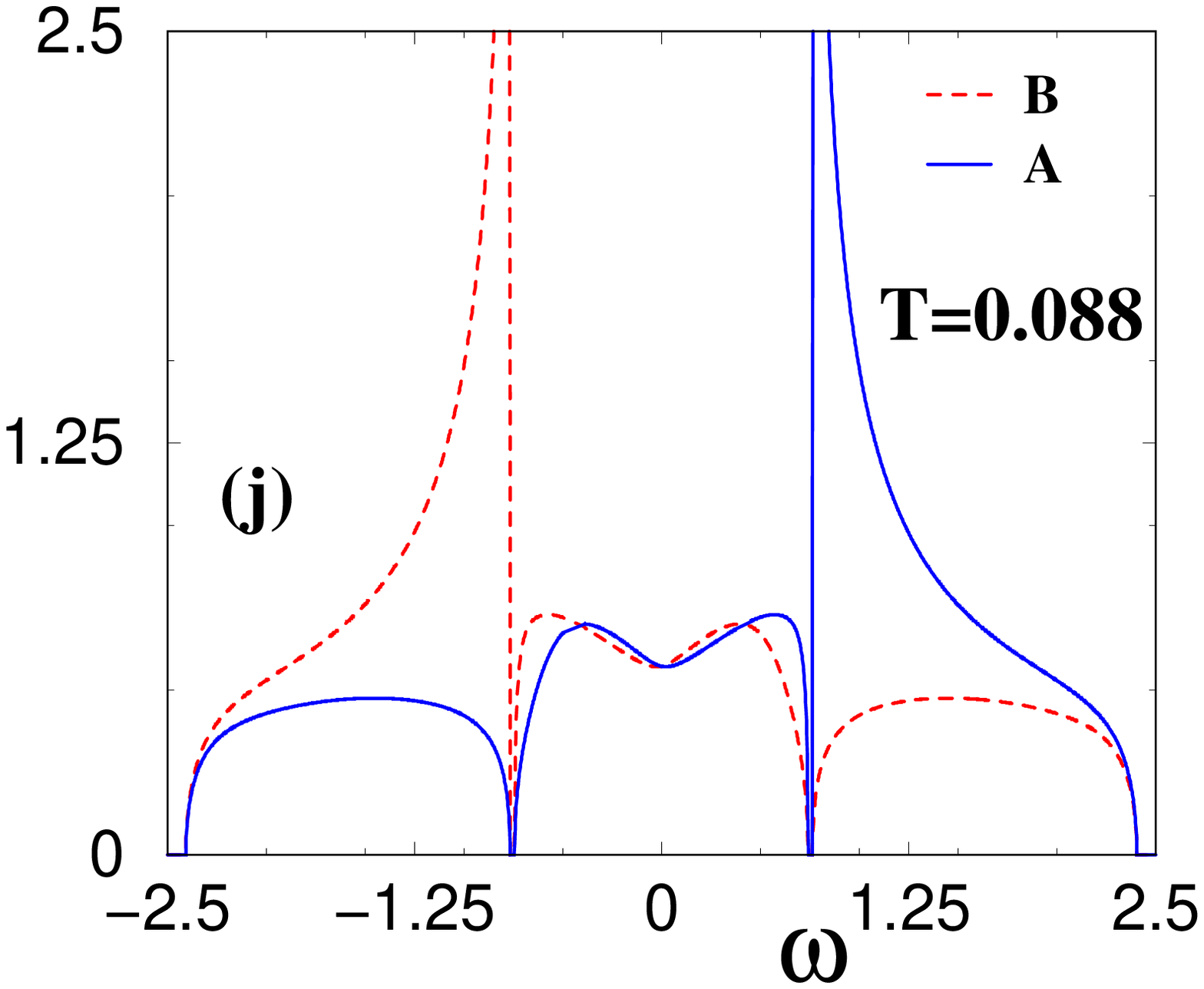}
\includegraphics[scale=0.34]{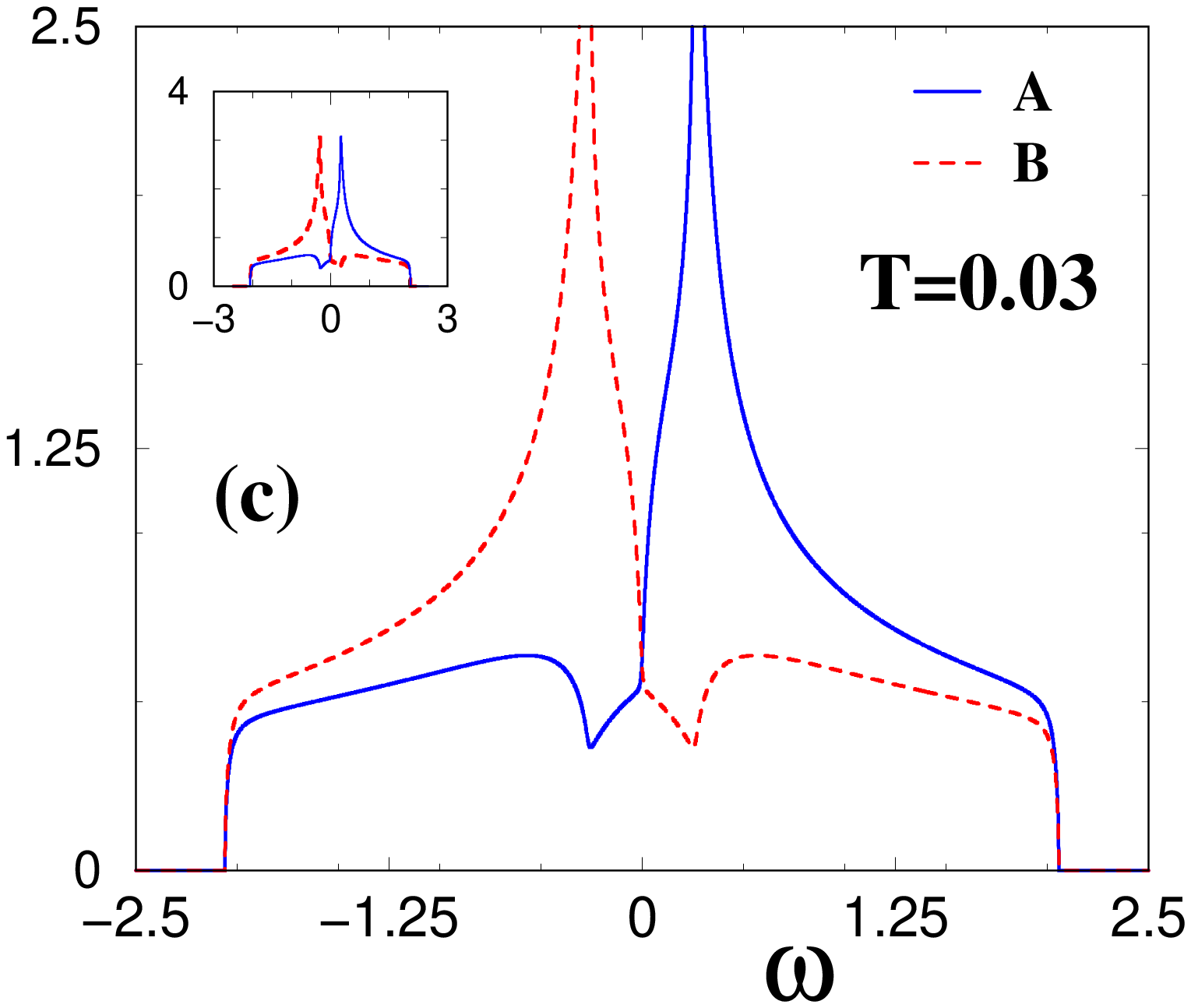}
\includegraphics[scale=0.34]{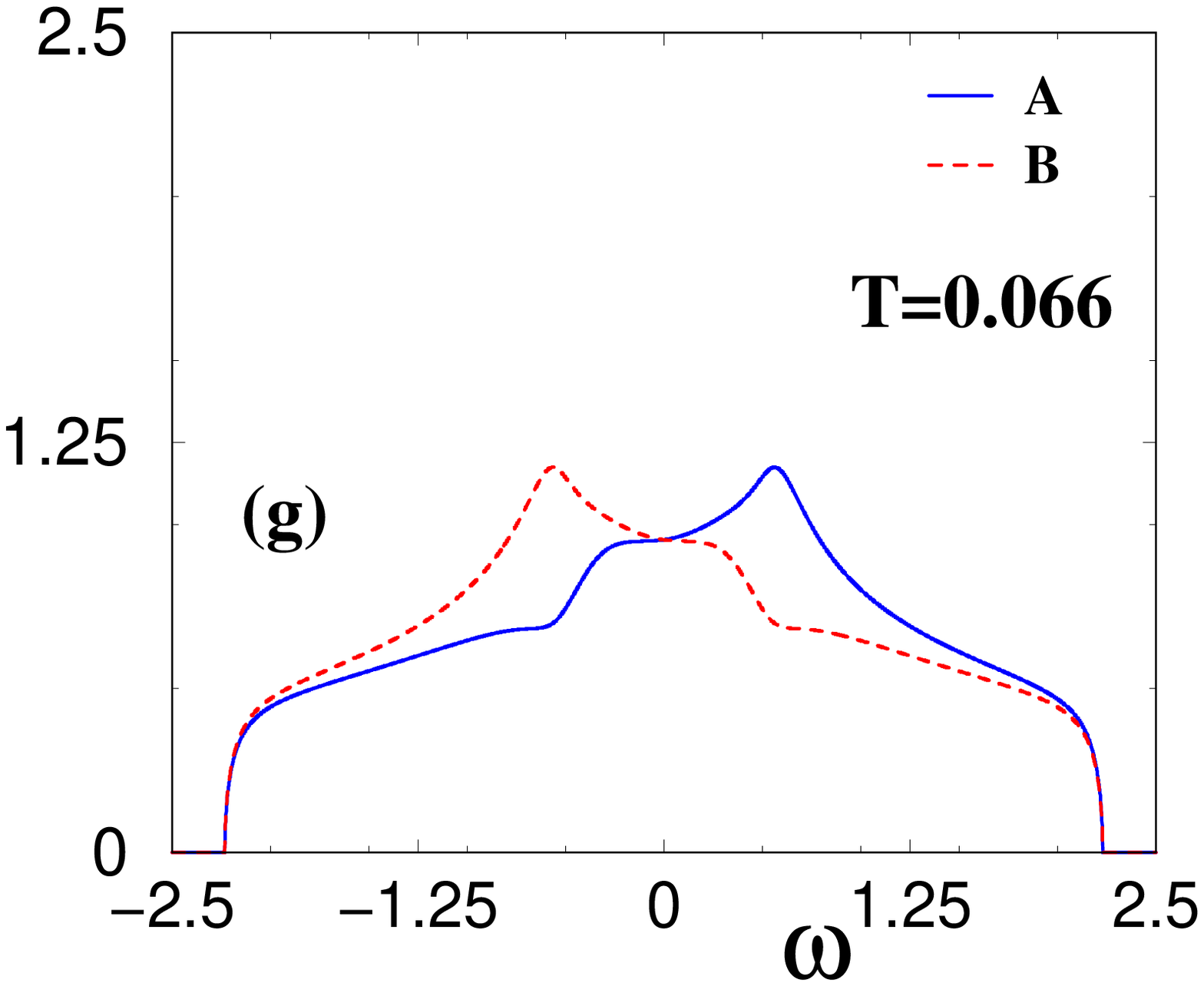}
\includegraphics[scale=0.34]{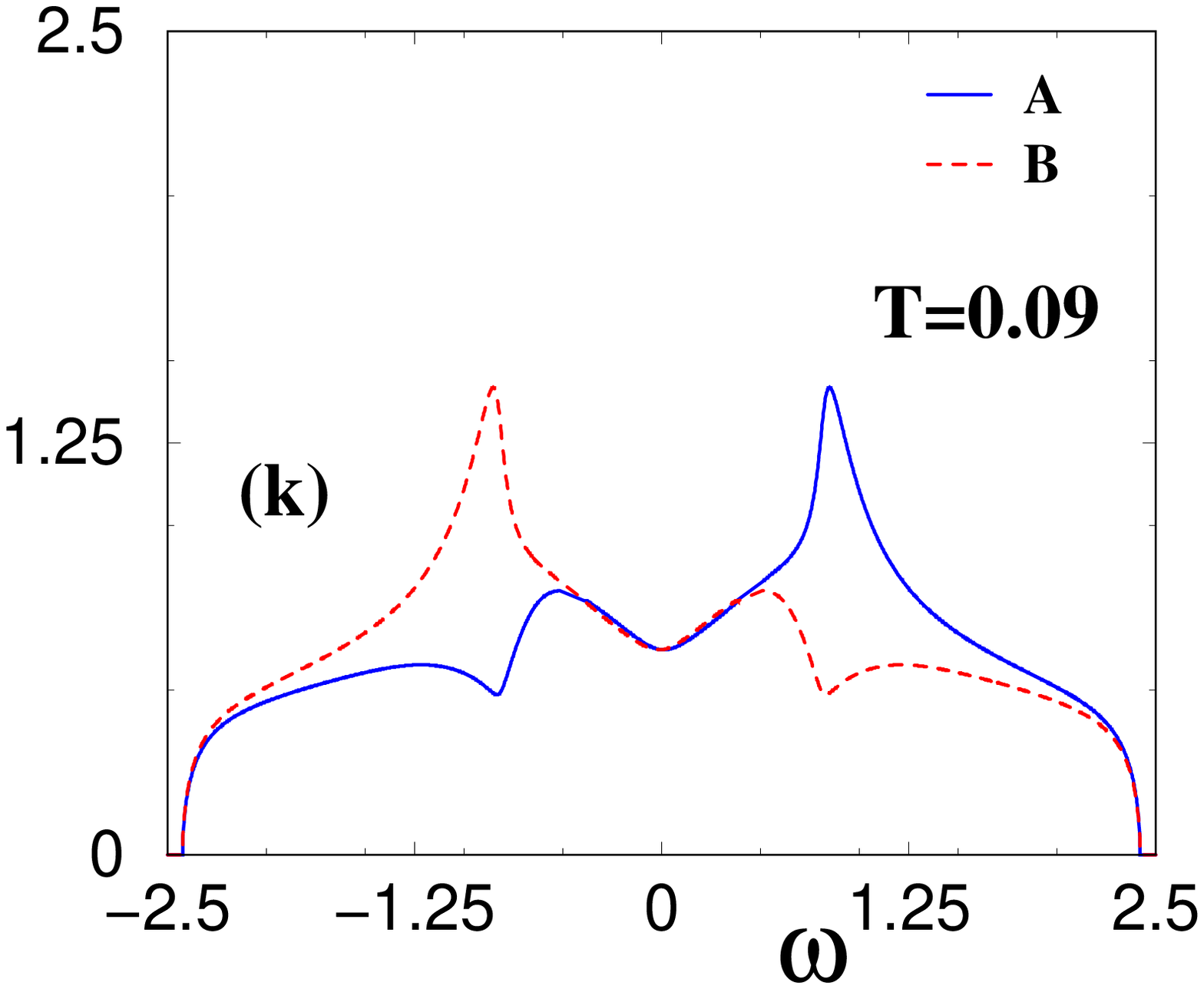}
\includegraphics[scale=0.34]{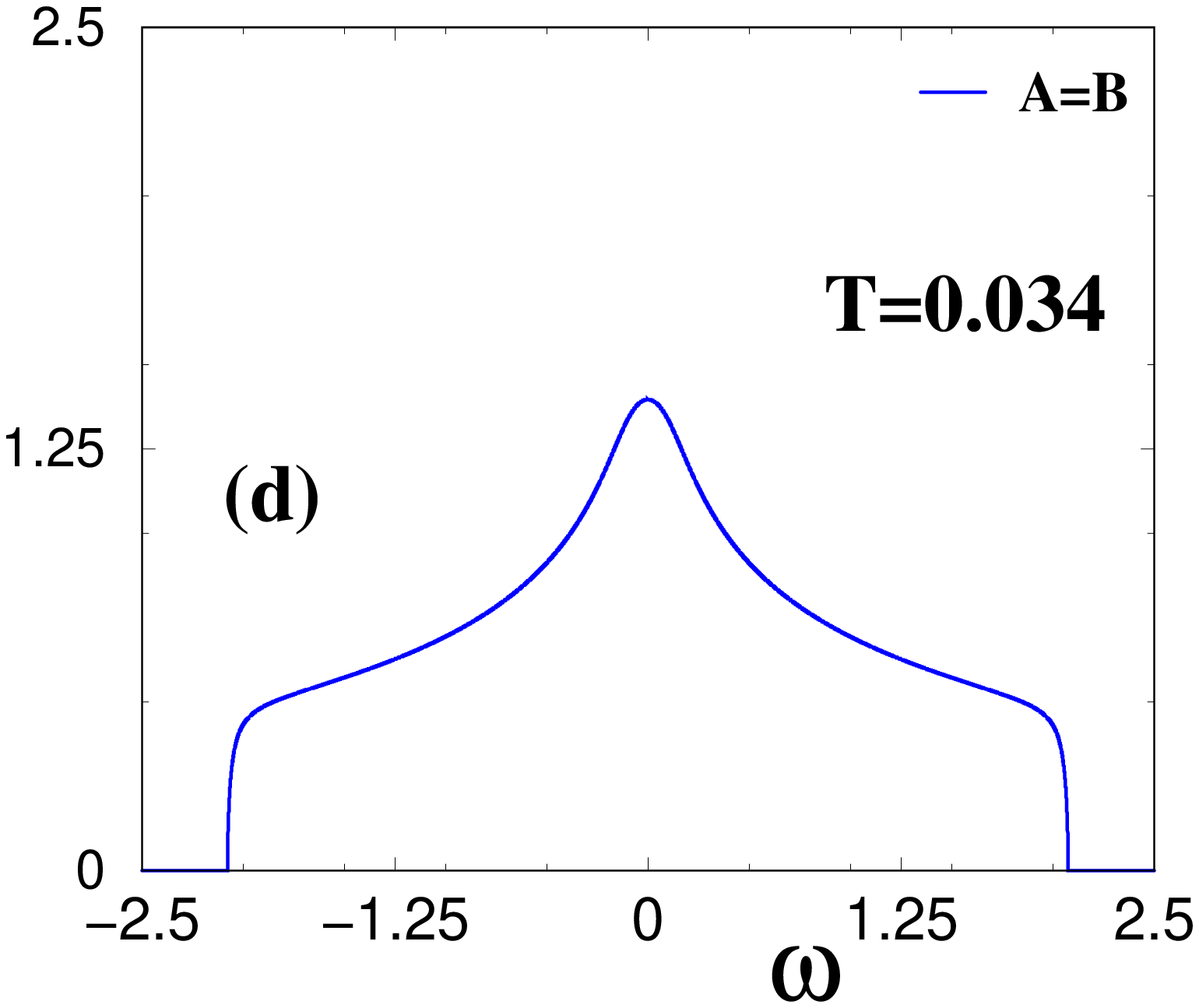}
\includegraphics[scale=0.34]{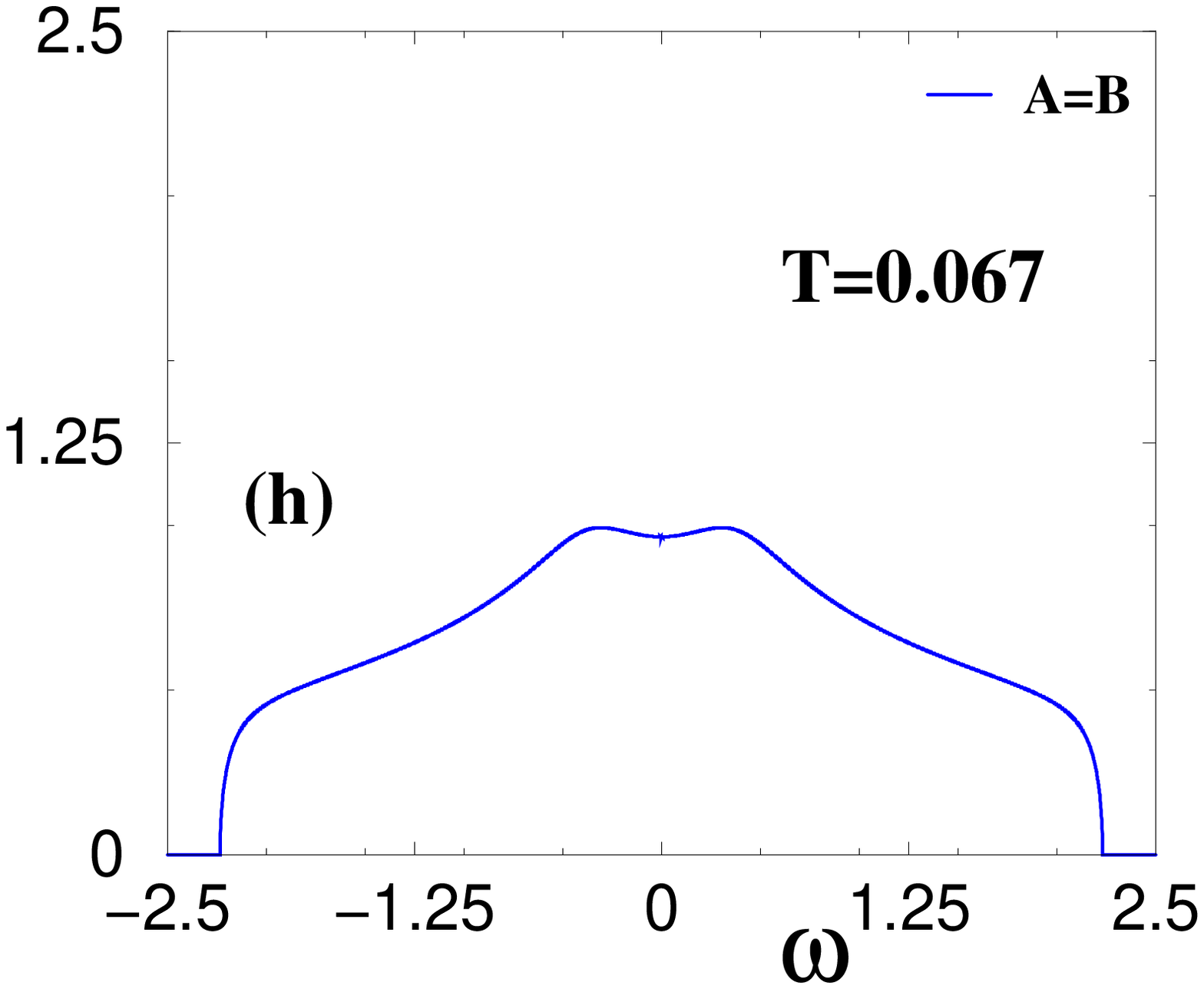}
\includegraphics[scale=0.34]{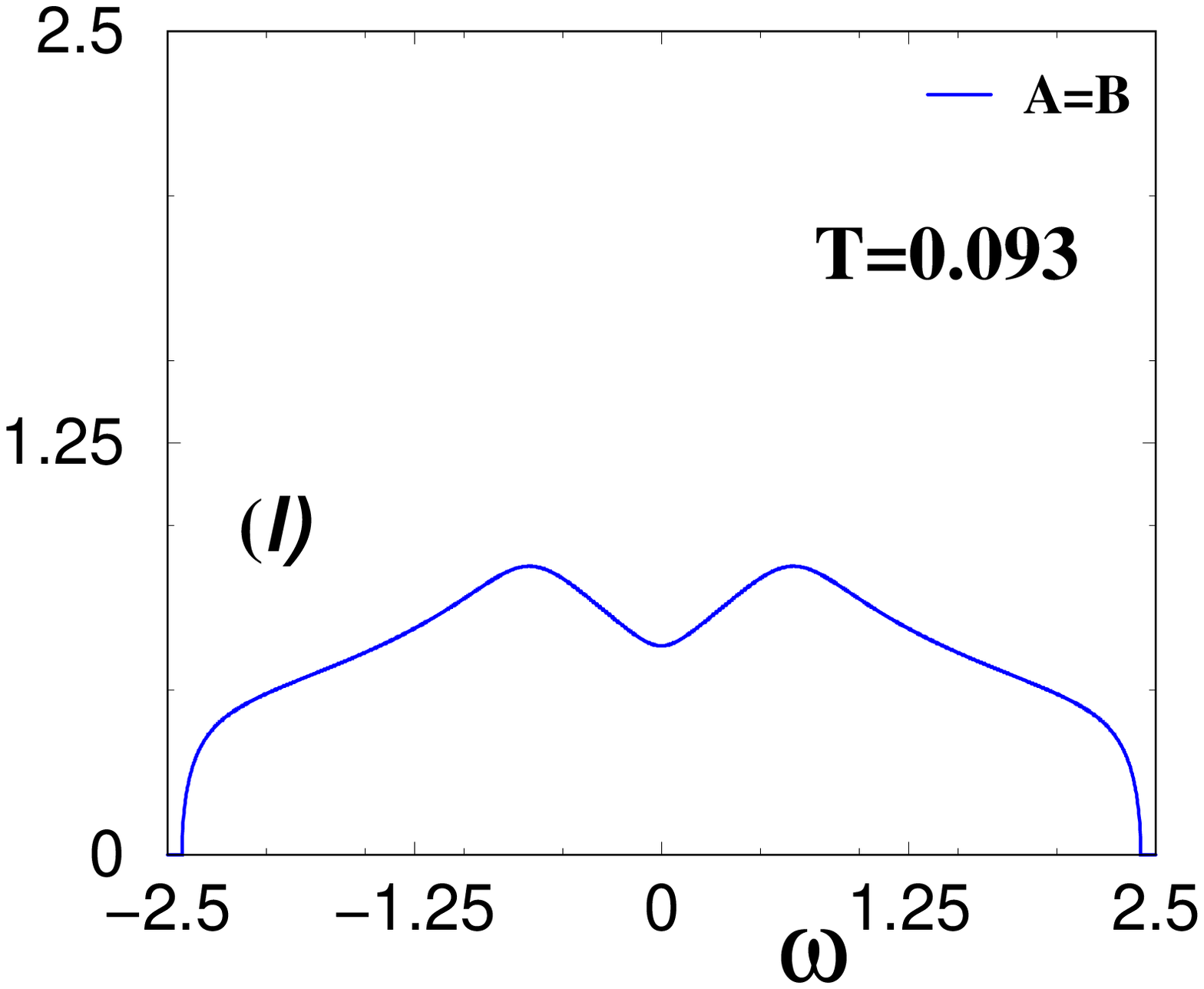}
\caption{(Color online) Diagrams showing the evolution of local spectral function
for 2DDOS as function of temperature for three values of
$U=0.5,1.0,1.5$ in the first, second, and third columns
respectively, and in increasing order of temperature. Dotted lines
are for sublattice $A$ and long dashed lines are for sublattice
$B$.} \label{figrdos}
\end{center}
\end{figure*}

This emerges from a study of results such as the ones shown in
Fig.s~\ref{chap2figb}-\ref{figrdos}. Fig.~\ref{chap2figb} shows the
CDW order parameter $(\Delta$w ) as a function of the temperature
$T$ for five values of $U=0.5, 1.0, 1.5, 2.0, 2.5$. In
Figs.~\ref{figsdos} and \ref{figrdos} we have shown the evolution of
the local spectral function for SDOS and 2DDOS \cite{localapproximation} respectively as a
function of temperature and $U$. In the uniform phase, the spectral
functions are temperature independent as we have remarked before,
and their $U$ dependence can be seen in the bottom panels of
Fig.~\ref{figsdos} for the SDOS case and in Fig.~\ref{figrdos} for
2DDOS \cite{localapproximation}. Basically, as U increases, there is a transfer of spectral
weight from low frequencies out to large frequencies, leading to the
formation of `Hubbard bands' and a `pseudo gap'. Eventually, for $U
> U_{MI}$ ($U_{MI} = 2$ for SDOS) a real gap opens up, and a metal
insulator transition ensues, as indicated in Fig.~\ref{Newphase}. As
we decrease the temperature at a fixed $U$ within the checker-board
phase, there is spectral weight transfer from low frequencies to
near a threshold frequency corresponding to the HF gap. However, as
can be seen in the next (second from bottom) set of panels in
Figs.~\ref{figsdos} and \ref{figrdos}, a real gap in the spectral
function does not develop just after one crosses into the
checker-board phase as would be predicted, for example, by the HF
approximation. Rather it develops only below the lower critical
temperature $T_l(U)$ shown in Fig. \ref{Newphase}. Even more
interestingly, $T_l(U)$ has a reentrant feature as function of
$U$ (see Fig.~\ref{Newphase}).

We note that at $T=0$, $\Delta$w=0.5,  $\Sigma_{n}^{A} = U$ and
$\Sigma_{n}^{B} = 0 $. The DMFT spectral functions at $T=0$ are thus
identical to the HF spectral functions, corresponding to split bands
with dispersion $\pm \sqrt {(U/2)^2 + \epsilon_{\bf k}^2}$. The gap
in the spectral function is hence exactly equal to $U$. However, as
we increase the temperature the value of $\Delta$w deviates
slightly from $0.5$ and the spectral function for each sublattices
develops two peaks around $\omega=0$ and one observes three gaps as
can be seen in the topmost panels in Figs. \ref{figsdos} and
\ref{figrdos}. As we increase the temperature further, these two
peaks grow in intensity and the gaps away from $\omega=0$ close up
and only the gap around $\omega=0$ remains. Eventually, past the
lower critical temperature $T_l(U)$ mentioned above, even the
gap around $\omega=0$ vanishes, but the order parameter is still
non-zero, until $T$ crosses $T_c$ whence $\Delta$w becomes zero and
the local spectral function becomes uniform. This behavior is
generic for $U < U_{MI}$. For $U > U_{MI}$, the spectrum is always
gapped.

In the gapless CDW phase B, the $b$ electrons are in a state which
allows gapless excitations and coexist with $\ell$ electrons in a
gapped state. We can view the effective $\ell$ electron energy as
the centroid of the $\ell$ electron spectral function. The gap in
$\ell$ electrons spectral function is therefore definable as
$\epsilon^*_{{\ell}_{A}}-\epsilon^*_{{\ell}_{B}}$. The absence of a gap
in the $b$-electron spectral function arises from the same source
that causes the $b$-electrons in the uniform phase to acquire a
non-Fermi liquid character, namely the $\ell$-electrons which act as
a source of disorder scattering, leading to a finite life time of
the $b$ electrons at the Fermi energy. In the CDW phase close to
$T_c$ and for $U$ not too large, although there is long range
staggered order in $n_{\ell}$, there are large thermal fluctuations
in $n_{\ell}$ which again act as disorder scatterers for the
$b$-electrons, leading to a finite life-time at the Fermi surface
and to a gapless $b$-electron spectral function. This feature goes
away at low temperatures or for large U, as the order in $n_{\ell}$
gets stronger and the fluctuations get reduced.

\begin{figure}[tbp]
%\begin{center}
\includegraphics[scale=0.45]{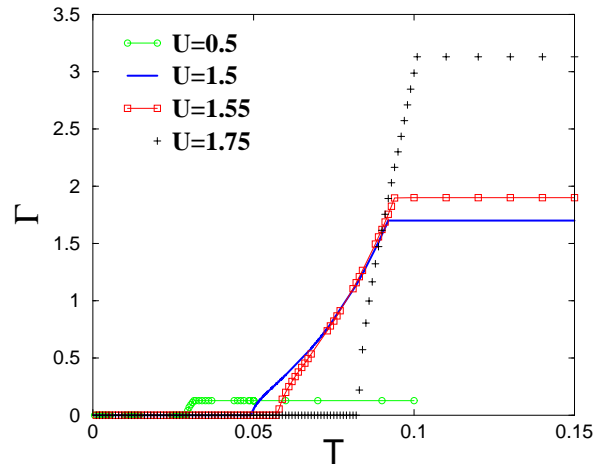}
%\end{center}
\caption{(Color online)Variation of $\Gamma (\equiv \Im \; (\Sigma_A + \Sigma_B)$
at $\omega = $0) with temperature for different values of $U$ for
SDOS.} \label{gamma1}
\end{figure}

\begin{figure}[tbp]
%\begin{center}
\includegraphics[scale=0.45]{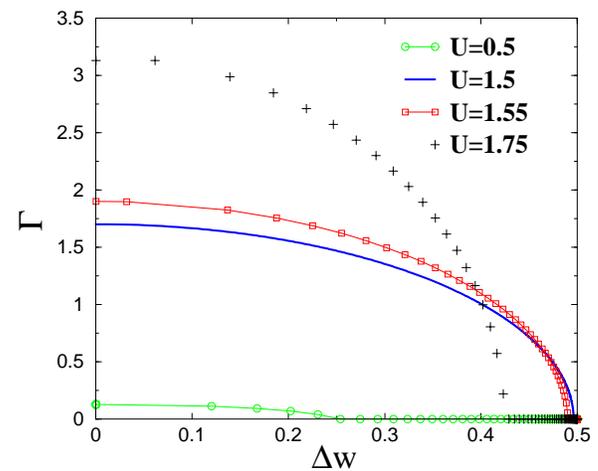}
%\end{center}
\caption{(Color online)Variation of $\Gamma$ with the order parameter ($\Delta$w)
for different values of $U$ for SDOS.} \label{gamma2}
\end{figure}

To get a quantitative estimate of the lower critical temperature
$T_l$, we have studied the quantity
$$\Gamma~\equiv~\Im~(\Sigma_{A}+~\Sigma_{B})_{\omega=0}$$ as a
function of temperature, as the vanishing of $\Gamma $ signals the
opening up of the gap around $\omega=0$. Figs. \ref{gamma1}  and
\ref{gamma2} depict $\Gamma $ as a function of temperature and
$\Delta w$ respectively for $U=0.5, 1.5, 1.55, 1.75$. We note that
at $U=0.5$ the gap closes up at $ T_l = 0.029$ (corresponding
$\Delta$w=0.240). As we increase $U$ the gap closes up at higher
values of temperature, {\it i.e.}, at $T_{l} = 0.052$ 
$(\Delta$w=0.435) for $U=1$. However, as we increase $U$ further, the gap
closes up at  lower values of $T_{l}$ up to $U$ very close to
$\sqrt{2}$ and $T_{l}=0.0108$ $(\Delta$w$\simeq 0.5)$. Beyond $U
\simeq \sqrt{2}$ the lower critical temperature $T_{l}$ for the
closing of the gap again increases to join the $T_c$ versus $U$
curves at $U=2.0$~. Hence we get the reentrant curve for $T_{l}$
as shown in Fig.~\ref{Newphase}.

\begin{figure}[tbp]
%\begin{center}
\includegraphics[scale=0.45]{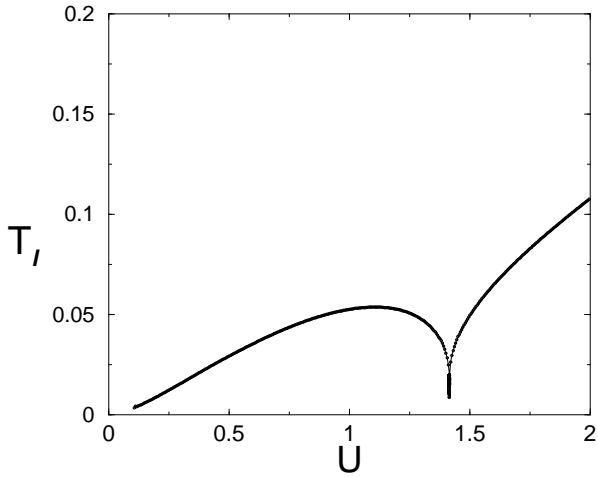}
%\end{center}
\caption{Variation of the critical temperature $T_{l}$ with U.}
\label{tl-vs-u}
\end{figure}

\begin{figure}[tbp]
%\begin{center}
\includegraphics[scale=0.45]{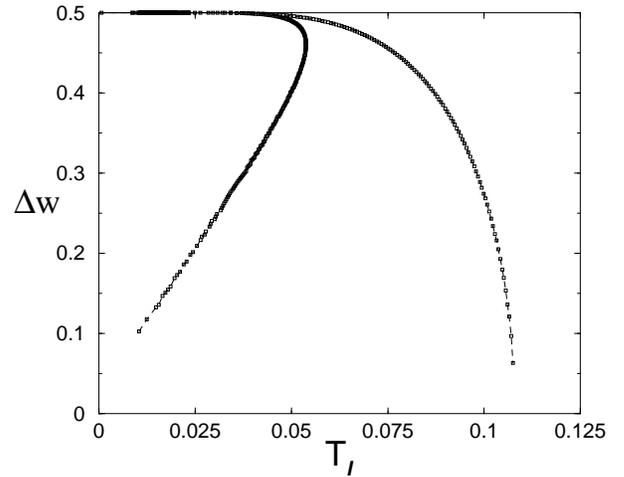}
%\end{center}
\caption{Variation of the order parameter ($\Delta$w) with the
critical temperature $T_{l}$.} \label{dw-vs-tl}
\end{figure}

\begin{figure}[tbp]
%\begin{center}
\includegraphics[scale=0.45]{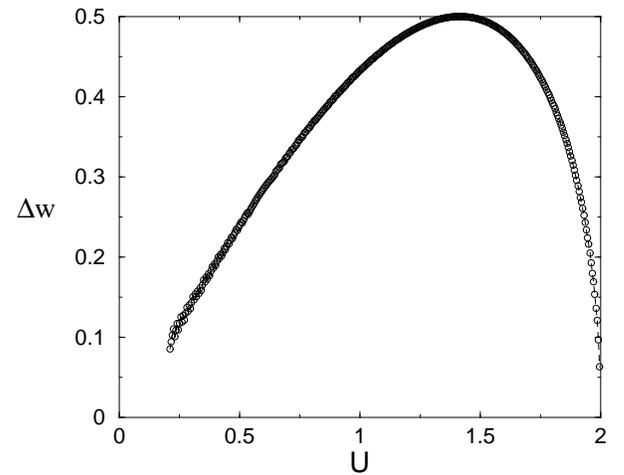}
%\end{center}
\caption{Variation of the order parameter ($\Delta$w) with U on the
curve $T_{l}$.} \label{dw-vs-u}
\end{figure}

The lower critical temperature $T_{l}$ shown in
Fig.~\ref{Newphase} was obtained by analyzing the data of spectral
functions and the order parameter as discussed above. Using the
relation $\Sigma^{A}(\omega=0)=-(\Sigma^{B}(\omega=0))^*$ and the
expression for $\Sigma^{A}(\omega=0)$ from
eq. (\ref{sigab}), and invoking the condition for the vanishing of
$\Gamma$, we obtain the following implicit equation for $T_{l}$:

\begin{equation}
\Delta \mbox{w}(T_{l}) + \frac{U}{2}Re{(G^{A}(\omega=0
;T_{l}))}=0. \label{eq:Tl}
\end{equation}
$T_{l}(U)$ can be very accurately determined using eq.
(\ref{eq:Tl}) and the results are exhibited in Fig.\ref{tl-vs-u},
and in the inset of Fig.\ref{Newphase}.  We have also computed the
order parameter ($\Delta$w) as a function of $T_{l}$ and U on
the curve $T_{l}(U)$, and we have shown these in
Figs.\ref{dw-vs-tl} and \ref{dw-vs-u}. As we move along the curve
$T_{l}(U)$ in Fig.~\ref{Newphase} the order parameter ($\Delta$w)
first increases, helped both  by the increase of $U$ and the
decrease of $T$, approaching a maximum value very close to 0.5 at U$
\simeq \sqrt{2}$. However, as we move to larger values of $U$, the
order parameter ($\Delta$w) starts decreasing and vanishes at U=2.0,
where $T_{l}$ joins $T_c$.

When $\Delta$w is very close to 0.5, then $\Sigma^{A}$ and
$\Sigma^{B}$ are very close to U and 0 respectively. Substituting
these values in eqs. (\ref{GAB}) and  (\ref{eq:Tl}), we obtain
\begin{equation}
Re\{(G^{A})^{-1}\} \simeq
-\frac{\frac{U}{2}+\sqrt{(\frac{U}{2})^2+4t^2}}{2} \simeq -U.
\label{eq:dip}
\end{equation}

Solving eq. (\ref{eq:dip}), we can verify that the dip of the lower
critical temperature $T_{l}$ occurs at  $U \simeq \sqrt{2}$.

\begin{figure}[tbp]
%\begin{center}
\includegraphics[scale=0.5]{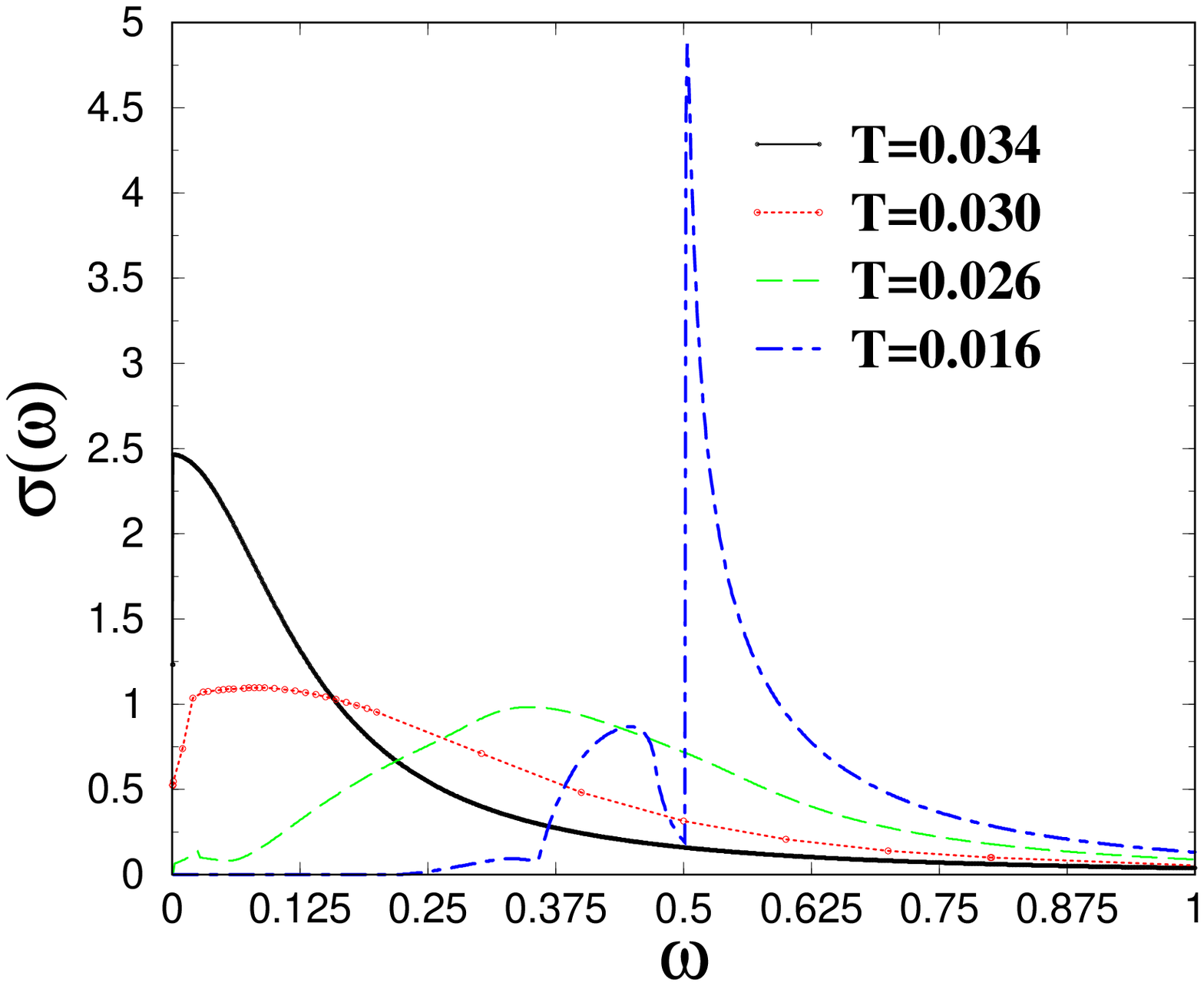}
%\end{center}
 \caption{(Color online) Optical Conductivity
$\sigma(\omega)$ at different temperatures for a fixed value of $U =
$ 0.5 for SDOS} \label{opsdU0.5}
%\begin{center}
\includegraphics[scale=0.5]{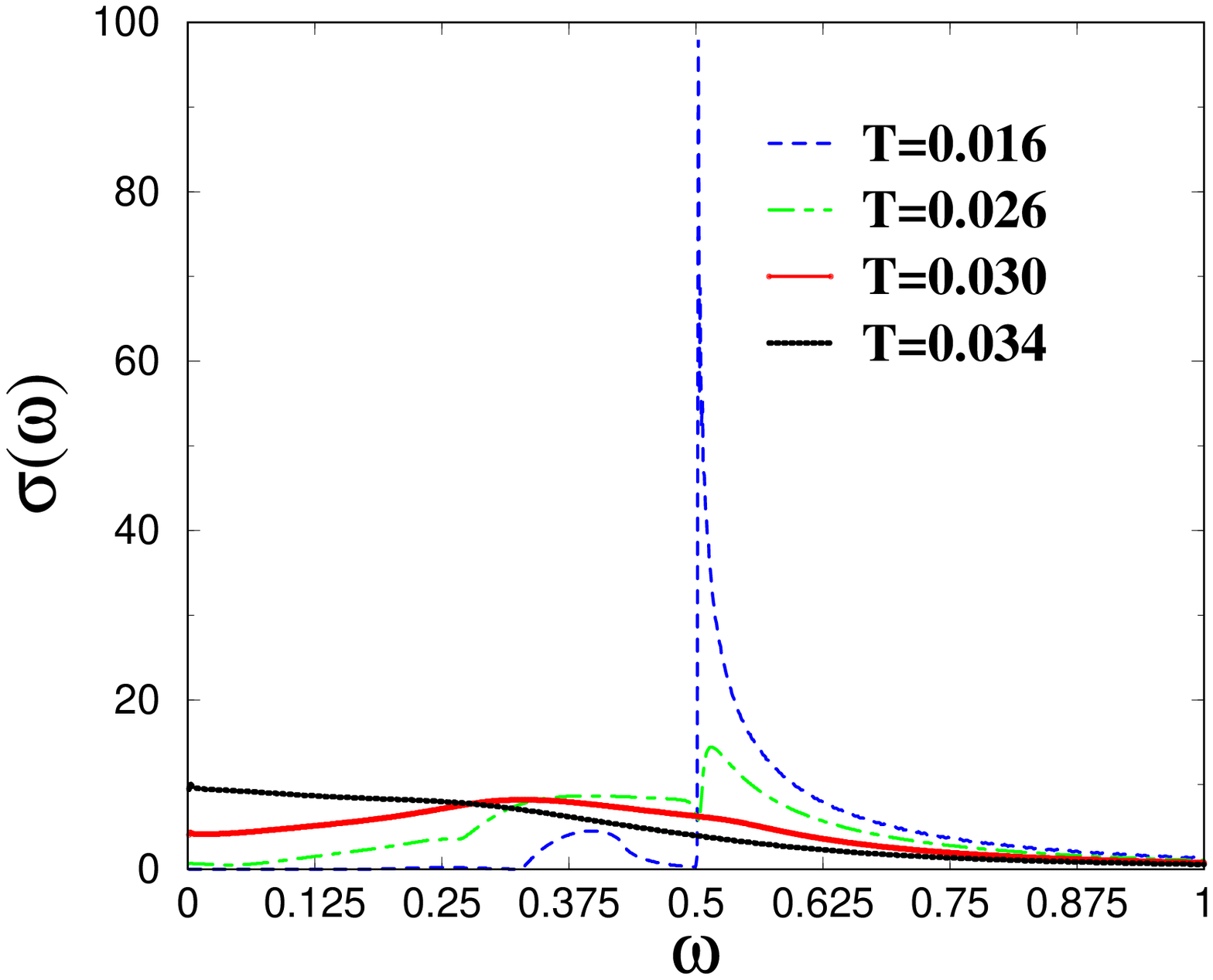}
%\end{center}
\caption{(Color online) Optical Conductivity $\sigma(\omega)$ at different
temperatures for a fixed value of $U= $ 0.5 for 2DDOS}
\label{oprdU0.5}
\end{figure}

\begin{figure}[tbp]
%\begin{center}
\includegraphics[scale=0.5]{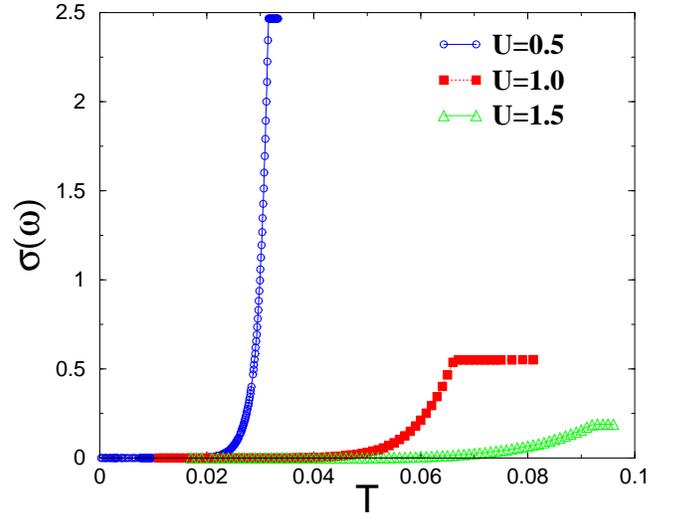}
%\end{center}

\caption{(Color online)dc-conductivity as a function of temperature for a fixed value
of $U = $ 0.5, 1.0, 1.5.}
\label{dccond}
\end{figure}
It is interesting to ask how these novel features in the spectral
functions affect experimentally measurable quantities. In
particular, we have studied the optical conductivity
$\sigma(\omega)$ as a function of T and U. In the CDW phase it can
be shown to be given by\cite{cox_2}
\begin{eqnarray}
\sigma (\omega)&=& \pi\frac{e^2}{2 \hbar a}
\int^\infty_{-\infty}d{\epsilon}D_{tr}(\epsilon)
 \int_{-\infty}^{\infty}\frac{d\omega}{2\pi} Tr({\sigma^{x}\bf
A}(\epsilon,\omega^\prime) \nonumber \\
&& \times {\sigma^{x}\bf A}(\epsilon,{\omega^\prime+\omega}))
\frac{n^{-}_F{(\omega^\prime)}-n^{-}_F{(\omega^\prime+\omega)}}{\omega}\mbox{,}
\label{eqnm}
\end{eqnarray}
where $\sigma^x $ is the Pauli matrix, ${\bf
A}(\epsilon,\omega^\prime) = -\frac{1}{\pi} \Im {\bf
G}_{\epsilon}(\omega^\prime)$, is the matrix spectral function for
the matrix Green's function given by eq. (\ref{latticeG}), and
$D_{tr}(\epsilon)$ is the transport DOS\cite{cox_2,hassan_2}. We
plot $\sigma(\omega)$ for different values of temperatures for a
fixed value of $U=0.5$ in Figs.~\ref{opsdU0.5} and \ref{oprdU0.5}
for the SDOS and 2DDOS respectively. When T=.016 one can see from
the spectral function (first diagram of column 1 of
Fig.~\ref{figsdos}) that the gap around $\omega=0$ is roughly
$0.25$, and there are two bands of low spectral weights which are
separated from the main bands by a second smaller gap $(0.05)$
around $\omega=0.5$. These features are reflected in
$\sigma(\omega)$ (see Fig.~\ref{opsdU0.5} dot-dashed curve) which
shows that there  is no optical response up to $\omega=0.25$. Then
there is a rise, a small dip around $\omega=0.5$, followed by a
sharp peak. As we increase the temperature to $T=0.026$ (see second
diagram of column 2 of Fig.~\ref{figsdos}), the spectral function
now has a very small gap around $\omega=0$, this feature of spectral
function is reflected in the corresponding $\sigma(\omega)$ (see
Figs.~\ref{opsdU0.5}(long dashed curve) and \ref{oprdU0.5}
(dot-dashed curve)). Similarly, the other two curves for the optical
conductivity at $T=0.03$ and $T=0.032$ (Fig.~\ref{opsdU0.5} dotted
and solid lines respectively) correspond to the spectral functions
in the last two diagrams of column 1 in Fig.~\ref{figsdos}.
We note however that the dc conductivity $\sigma(\omega=0) $ does
not capture these features, and reflects only the transition from
the CDW to the uniform phase which shows up as a slope discontinuity
at $ T_c $, as can be seen in Fig.~\ref{dccond}.

\section{Concluding Discussion}

In conclusion, we have presented results from a detailed DMFT study
of the spectral functions in the CDW phase of the half-filled SFKM
as function of temperature and U.  We have shown that the proximity
of the non-Fermi liquid metallic phase affects the CDW phase,
leading to a region in the phase diagram where we a get CDW phase
{\it without a gap in the spectral function}. Interestingly, this
gapless CDW phase shows a {\it reentrant transition} to the gapped
CDW phase as $U$ increases. This is a radical deviation from
mean-field prediction where the CDW phase is always gapped, with the
gap being proportional to the order parameter. We have also
discussed how these features affect response functions, {\it e.g.},
the optical conductivity. It would be interesting to study whether,
and to what extent these features survive when one goes beyond DMFT
for the SFKM, {\it e.g.} in more sophisticated approximations such
as the Dynamical Cluster Approximation\cite{HRK_1,HRK_2},
cluster-DMFT \cite{c-dmft} or Variational Cluster Approximation
\cite{VCA}, which include the effects of short range inter-site
correlations.

We note that gapless CDW phases are easy to achieve even within the
Hartree approximation by considering second neighbor hopping, which
gets rid of nesting. What is novel about the gapless CDW phase
discussed here is that it is correlation induced, and appears
despite the presence of perfect nesting. It is interesting to ask
whether such phases can appear in other models with strong
correlations. The normal (repulsive) Hubbard model does not have CDW
instabilities. But an extended Hubbard model with nearest neighbor
repulsion $V$ would. We believe that at intermediate values of the
Hubbard U and appropriate values of $V$ the extended Hubbard model
with nearest neighbor hopping could exhibit a gapless commensurate
CDW phase. For, the DMFT treatment of such a model in the presence
of a CDW would correspond closely with the recent study\cite{garg}
of correlation effects in a two sublattice band insulator where
correlations were shown to induce metallicity.

On the experimental front, gapless CDW phases have recently been
observed in 2H- transition metal dichalcogenides (eg., ref.
\onlinecite{2H-TMDs} and references therein). In these compounds,
strongly coupled electronic and lattice degrees of freedom are
involved in the generation of the CDW instability, and the gapless
feature has been attributed to very large second neighbor hopping.
Since in our model we are considering only electronic degrees of
freedom, and only nearest neighbor hopping, we have not compared the
experiments directly with our findings.

\section{Acknowledgments}

SRH thanks the Council of Scientific and Industrial Research(India)
and NSERC (Canada) for financial support, and gratefully
acknowledges useful discussion with G.Venkateshwara Pai and R. Karan.
HRK's research was supported by the University Grants Commission
(India), the Department of Science and Technology(India) and the
Indo-French Centre for the Promotion of Advanced Research (grant no.
2400-1). He would also like to acknowledge the hospitality of the
KITP (supported by NSF grant no. PHY05-51164) during the preparation
of the revised manuscript.

%%%%%%%%%%%%%REFERENCES%%%%%%%%%%%

%\end{multicols}

\begin{references}
\bibitem{Footnote}
The Bethe lattice is a connected cycle-free graph where each node is
connected to $Z$ neighbors, where $Z$ is called the coordination
number. This leads to a tree-like structure emanating from a central
node (called the root or origin of the lattice), with all the nodes
arranged in shells around the central one. The number of nodes in
the kth shell is given by $N_{k} = Z{(Z-1)}^{(k-1)}$ for $k
> 0$. A checkerboard phase on the Bethe lattice corresponds to
placing $b$ and $\ell$ electrons on alternating nodes.
\end{references}
\end{document}